\documentclass[preprint,review,12pt,authoryear]{elsarticle}
\usepackage{amssymb, amsmath, psfrag, graphicx,color}
\usepackage{units,booktabs,url,makecell,xcolor,multirow, diagbox}

\def\ve#1{\mathchoice{\mbox{\boldmath$\displaystyle#1$}}
{\mbox{\boldmath$\textstyle#1$}}
{\mbox{\boldmath$\scriptstyle#1$}}
{\mbox{\boldmath$\scriptscriptstyle#1$}}}

\newcommand{\eg}{\text{e.g., }}
\newcommand{\ie}{\text{i.e., }}
\newcommand{\wrt}{\text{w.r.t. }}

\journal{Computers \& Chemical Engineering}

\begin{document}
\begin{frontmatter}
\title{Robust multi-stage model-based design of optimal experiments for nonlinear estimation}
\author[TUDO]{Anwesh Reddy Gottu Mukkula}
\author[STUBA]{Michal Mate\'a\v s}
\author[STUBA]{Miroslav Fikar}
\author[STUBA,cor1]{Radoslav Paulen}
\cortext[cor1]{Tel.: +421 (0)2 5932 5730, Fax: +421 (0)2 5932 5340 (R. Paulen)}
\address[TUDO]{Process Dynamics and Operations Group, Technische Universit\"at Dortmund, Emil-Figge-Strasse~70, Dortmund 44227, Germany}
\address[STUBA]{Faculty of Chemical and Food Technology, Slovak University of Technology in Bratislava, Radlinsk\'eho 9, 812 37 Bratislava, Slovakia}

\begin{abstract}
We study approaches to the robust model-based design of experiments in the context of maximum-likelihood estimation. These approaches provide robustification of model-based methodologies for the design of optimal experiments by accounting for the effect of the parametric uncertainty. We study the problem of robust optimal design of experiments in the framework of nonlinear least-squares parameter estimation using linearized confidence regions. We investigate several well-known robustification frameworks in this respect and propose a novel methodology based on multi-stage robust optimization. The proposed methodology aims at problems, where the experiments are designed sequentially with a possibility of re-estimation in-between the experiments. The multi-stage formalism aids in identifying experiments that are better conducted in the early phase of experimentation, where parameter knowledge is poor. We demonstrate the findings and effectiveness of the proposed methodology using four case studies of varying complexity.
\end{abstract}

\end{frontmatter}

\section{Introduction}
The process systems engineering community adopts mathematical models successfully in various endeavors such as product and plant design, control system design, operations optimization, etc.~\citep{pan13, fun16, saf16}. A mathematical model is usually an abstract representation of a true system via sets of equations (algebraic, ordinary differential, or partial differential), inequalities (e.g., a range of model validity), and logical conditions. 

A crucial part of model development is the design of experiments (DoE). DoE is a branch of mathematics that uses the nature of the model being built to determine the most favorable experimental conditions~\citep{fis35, goodwin1977dynamic, box78, esp89, hja05, fra08, pro08, wol08, pro13, fed13}. Such conditions allow for the most informative data to be collected during the experiment, implying the highest quality of the fitted model. In linear estimation, the model behavior is linearly proportional to the values of the parameters, and thus the actual values of parameters do not play a major role in a successful experiment design. The situation is different in nonlinear estimation, where the model nature is highly influenced by its parameter values. Such a phenomenon gives rise to striking differences between linear and nonlinear experiment design and---besides the increased numerical complexity of the problem~\citep{kor04, sch09}---it makes the result of nonlinear experiment design to be heavily dependent on the a priori knowledge of the estimated parameters~\citep{asp02, roj07}. This is also commonly referred to as a chicken-and-egg problem, where in order to estimate parameters with high precision, one requires very precise knowledge of the parameters a priori.

A standard procedure of the experiment design for nonlinear models is to select a nominal value of the estimated parameters and to design the experiment for the nominal model. Naturally, if the nominal parameter values are far from reality, the nominal design will be significantly suboptimal or even infeasible to conduct in the presence of strict plant constraints~\citep{pronzato1985robust}. Several ways have been proposed to circumvent this problem such as robust min-max optimization~\citep{walter1987optimal, pronzato1988robust, mar06} and  stochastic~\citep{pronzato1985robust, gal10, str14, mes15, nim20} or scenario-based~\citep{Telen_roed, wel09} approaches. A good overview of the different approaches, which we will for simplicity denote as approaches to the robust design of experiments (rDoE), is available in~\cite{asp02, roj07}. A majority of these approaches concentrates primarily on fulfilling the operational constraints of the plants when running the experiments. Secondly, they try to recover the optimality or decrease the optimality loss by optimizing the experiments for the worst-case or the most probable model, given some a priori knowledge.

There are often situations in the experimentation phase of the model building where one can perform a series of successive experiments and is aware of the fact that the initial experiment run (designed using a nominal model) might not be optimal~\citep{van21}. This case also includes running robustly designed experiments presented in the previous paragraph. At the same time, one can learn from the previous experiments such that the next experiment is designed better---based on the model parameters re-estimated from the newly available data. This gives rise to the design of sequential experiments~\citep{bar_seqoed, seb00, olo19, pau19}. To the best of the authors' knowledge, an approach to DoE, where one optimizes a whole series of experiments with explicit accounting for the possibility of re-designing the successive experiments once new information is available, has not been pursued in the design-of-experiments literature despite that \eg physicians envision systematic approaches of this type (two- or multi-stage experiment design) for years~\citep{sim89, sve19}.

In this paper, we study the problem of rDoE, we assess the performance of different aforementioned techniques, and we propose a new rDoE technique for the situation, where one plans several consecutive experiments ahead while taking explicitly into account the possibility of re-estimating the parameters in-between the experiments. A mathematical tool for this task is a well-known two-stage or multi-stage programming~\citep{Garstka1974}, which is combined with our previous works on the robust experiment design in the context of set-membership estimation~\citep{anwesh17esc}. In this context, we study the experiment design using standard linearized design criteria---i.e., criteria based on the Fisher information matrix. We dedicate this paper to the proof-of-concept and the comparison with the most known approaches to the robust experiment design. Thus we concentrate on the simplest types of the experiment design problems for nonlinear estimation yet we provide suggestions on tackling more involved problem types at appropriate places in this paper.

We organized the paper as follows. The concepts of parameter estimation are introduced first. Next, the formulation of the experiment design is reviewed using linearized confidence regions. Further, the most well-known rDoE approaches are briefly introduced and the proposal for robust multi-stage DoE is outlined in detail. The case study section is then devoted to the numerical implementation and to an illustration and analysis via simulation examples using four simple nonlinear models.

\section{Preliminaries}
\subsection{Mathematical Model}
In this work, we consider that a mathematical model of a system with $n_p$ uncertain parameters $\ve p$, $n_u$ manipulated variables $\ve u_\tau$, and $n_y$ output variables $\hat{\ve y}$ is represented as
\begin{equation}
  \label{eq:model_static}
    \quad \hat{\ve y}(\ve p, \tau) = \ve F(\ve p, \ve u_\tau),
\end{equation}
where $\tau$ represents an ordinal number of a data point from one or more experiments. We assume that the mapping function $\ve F:\mathbb{R}^{n_p}\times\mathbb{R}^{n_u}\rightarrow\mathbb{R}^{n_y}$ is continuous and at least twice differentiable. Furthermore, we assume that the mapping function $\ve F$ is structurally correct and all the uncertain parameters $\ve p$ are identifiable. Therefore, the uncertain parameters $\ve p$ could be estimated upon having $N$ instances of plant measurements $\ve y_m$, with each instance consisting of $n_y$ measured variables, which could be obtained by performing experiments. It is also assumed that the plant measurements $\ve y_m$ are corrupted with white Gaussian noise.

Throughout this paper, we consider that the mathematical model~\eqref{eq:model_static} is static and explicit \wrt the output variables. Nevertheless, we note that the proposed methodologies can be extended straightforwardly to dynamic and implicit models. In the following subsections, existing methods are briefly presented for the identification of the uncertain parameters $\ve p$ and their respective confidence regions.

\subsection{Parameter Estimation}\label{sec:lse}
\label{subsec:Linear_Parameter_Estimation}
Under the assumption of uncorrelated and normally distributed measurement noise with a \emph{known standard deviation} vector $\ve\sigma:=(\sigma_1, \sigma_2,\hdots,\sigma_{n_y})^T$, the maximum-likelihood estimate is found via the least-squares estimation as
\begin{align}
\hat{\ve p} &= \arg\min_{\ve p} \sum_{i=1}^{n_y} \sum_{\tau=\tau_1}^{\tau_N} \sigma_i^{-2} (y_{m, i}(\tau) - \hat{{y}}_{i}(\ve{p},\tau))^2. \label{eq:w_with_variance}
\end{align}

The linearized (asymptotic) joint-confidence region of parameter estimates is then
given by an ellipsoid~\citep{SeberWild200309}
\begin{equation}\label{eq:LPE}
(\ve{p} - \hat{\ve{p}})^T \mathbb{FIM}(\hat{\ve p}, \mathcal U_N)
(\ve{p} - \hat{\ve{p}}) \leq \chi^2_{\alpha,n_p},
\end{equation}
where $\mathcal U_N:=(\ve u_{\tau_1}^T,\ve u_{\tau_2}^T,\dots,\ve u_{\tau_N}^T)^T$, $\chi^2_{\alpha, n_p}$ represents the upper $\alpha$ quantile of the chi-squared statistical distribution with $n_p$ degrees of freedom, and $\mathbb{FIM}$ is the so-called Fisher information matrix given by
\begin{equation}\label{eq:fim}
\mathbb{FIM}(\hat{\ve p}, \mathcal U_N) := \sum_{\tau=\tau_1}^{\tau_N} \mathbf Q(\hat{\ve p}, \ve u_\tau)^T \text{diag}^{-2}(\ve\sigma) \mathbf Q(\hat{\ve p}, \ve u_\tau),
\end{equation}
with the operator $\text{diag}(\ve a)$ signifying a diagonal matrix with vector $\ve a$ as the main diagonal and with the corresponding sensitivity matrix
\begin{equation}
 \mathbf Q(\hat{\ve p}, \ve u_\tau) := \frac{\partial \ve{F}(\ve{p},\ve{u}_\tau)}{\partial \ve p}\bigg|_{\hat{\ve p}}.
\end{equation}

We note here that if the variance of the measurement noise is unknown, it has to be estimated using experimental data. The formulations of the least-squares estimation problem and the joint-confidence region change correspondingly~\citep{fra08}---a quantile of F-distribution is used instead of $\chi^2$ distribution---yet they retain the same mathematical forms. The same claim holds for the case of correlated measurement noises. Thus---despite not explicitly involving these situations in our study---the obtained results here are directly applicable.

We also note that here---despite using standard asymptotic confidence regions in this paper---that there exist non-asymptotic confidence regions~\citep{cam05, per18}, which can be used in nonlinear parameter estimation and which can be considered in an extension of the present work.

\subsection{Model-based Design of Experiments}\label{sec:OED}
After stating the overall assumptions of this study, we recall here a methodology for DoE using the linearized confidence regions. We will consider---without loss to generality even for the latter rDoE---that the only constraints of the experiment are present by the limits of the experimental degrees of freedom $\mathcal U_N$. We make appropriate comments regarding the presence of constraints in the analyses below. We also assume that an estimate $\hat{\ve p}$---i.e., an expected value of $\ve p$---is available. The statistical properties of this assumption will be stated below.

Several design criteria are proposed in the literature~\citep{fra08} such as A, D, E, Modified E, V, Q, M, etc. Each of these designs aims to tune a particular property of the confidence region. We will focus our study on one of the most used criteria---i.e., the A design---yet we note that other design criteria might be considered as well using the ideas presented herein.

\begin{figure}
  \centering
    \psfrag{p1}[cc][cc][1]{$p_1$}
    \psfrag{p2}[cc][cc][1]{\ $p_2$}
    \centering\includegraphics[width=0.5\linewidth]{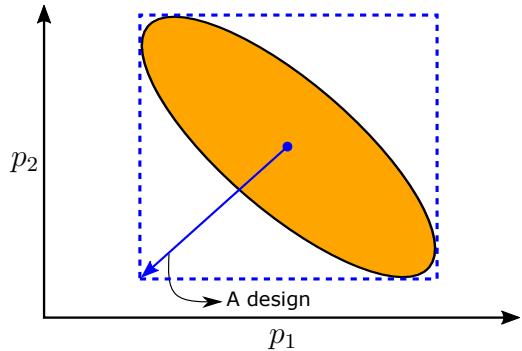}
    \caption{Illustration of the A design.}
\label{fig:oed_lin}
\end{figure}

The problem of designing A-optimal experiments can be formulated as an optimization problem of the form~\citep{fra08}
\begin{subequations}\label{eq:oed_lin}
\begin{align}
  \min_{\mathcal U_N:=(\ve u_{\tau_1}^T,\ve u_{\tau_2}^T,\dots,\ve u_{\tau_N}^T)^T} &\phi_A(\mathbb{FIM}(\hat{\ve p}, \mathcal U_N)):= \min_{\ve u} \text{trace}(\mathbb{FIM}^{-1}(\hat{\ve p}, \mathcal U_N))\\
  \text{s.t. } & \forall \tau \in\{\tau_1,\tau_2,\dots,\tau_N\}:\notag\\ \hat{\ve{y}}&(\hat{\ve p},\tau) = \ve{F}(\hat{\ve p}, \ve{u}_{\tau}),\\
  &\ve u^L\leq \ve u_\tau \leq \ve u^U.
\end{align}
\end{subequations}
This optimizes the experimental conditions such that the area of the box (the circumference in two dimensions) that encloses the confidence region is minimized. The bounds $\ve u^L$ and $\ve u^U$ represent the lower and upper limits of the experimental degrees of freedom. The idea is illustrated in Figure~\ref{fig:oed_lin} for a two-dimensional parametric space.

For the remainder of this paper, we will refer to the presented DoE method as \emph{nominal} design of experiments.

We note that the presented robustification techniques can also be readily used for a nonlinear DoE---i.e., the design of experiments using exact (nonlinear) confidence regions~\citep{rooney2001}. The design of experiment methodology in this context was recently studied in~\cite{mupa19_jpc}, \cite{WALZ201892}, and~\cite{wal20}.

\section{Robust Model-based Design of Experiments}
\label{sec:Robust_DoE}
If the nominal parameters $\hat{\ve p}$ are far from the true values $\ve p^*$, the nominal DoE might be significantly suboptimal. This section summarizes ways of robustifying the DoE with well-known rDoE techniques and proposes a methodology based on multi-stage decision making. We shall assume that a set $\ve P$ is available such that $\ve{p^*}\in\ve P$.

\subsection{Sequential Approach}\label{subsec:seq}
If the experimenter is able to run a series of $N_e$ experiments (adding up to the total of $N$ experiments), one of the simplest robustifying schemes is the sequential design~\citep{bar_seqoed}. Here one iteratively adjusts the value of $\hat{\ve p}$ based on the conducted experiments and re-conducts the nominal design (Section~\ref{sec:OED}). The following pseudo-algorithm can be used here:
\begin{enumerate}
 \item Set $\hat{\ve p}$ such that $\hat{\ve p}\in\ve P$.
 \item Perform the nominal DoE by solving problem~\eqref{eq:oed_lin} for $N_e<N$ experiments using the current $\hat{\ve p}$. \label{it:alg_seq_DoE}
 \item Apply the resulting $\mathcal U_{N_e}^*$ to the plant and obtain the measurements $\ve y_m$.
 \item Identify the new value of $\hat{\ve p}$ and the corresponding confidence region using the least-squares estimation (Section~\ref{sec:lse}) over all the past experiments.
 \item Go to step~\ref{it:alg_seq_DoE} unless:
 \begin{enumerate}
 \item the maximum (user-defined) number of possible experiments
 was conducted, or
 \item only a marginal (user-defined) improvement of the confidence region---measured by the chosen design criterion---was reached, or
 \item only a marginal (user-defined) update of the least-squares estimates was obtained.
 \end{enumerate}
 \item Terminate.
\end{enumerate}
Clearly, the performance of this approach is upper-bounded by the performance of the nominal design.

\subsection{Min-max Approach}\label{subsec:minmax}
Min-max formulation of DoE~\citep{pronzato1988robust} identifies the optimal experiment conditions $\mathcal U_N$ under the worst-case realization of $\hat{\ve p} \in \ve P$. The mathematical formulation of this problem reads as
\begin{subequations}\label{eq:minmax}
  \begin{align}
  \min_{\mathcal U_N}& \max_{\hat{\ve p}\in\ve P} \phi_A(\mathbb{FIM}(\hat{\ve p}, \mathcal U_N))\\
  \text{s.t. } & \forall \tau\in\{\tau_1,\tau_2,\dots,\tau_N\}:\notag\\  
  &\quad \hat{\ve y}(\hat{\ve p}, \tau)= \ve{F}(\hat{\ve p}, \ve{u}_{\tau}),\\
  &\quad \ve u^L\leq \ve u_\tau \leq \ve u^U.
  \end{align}
\end{subequations}
The bi-level nature of the presented problem might be simplified by appropriate sampling of the set $\ve P$. The approximate problem can be solved using epigraph reformulation. The samples of the set $\ve P$ can be selected as combinations of minimal and maximal parameter values from $\ve P$ if one assumes the effect of the uncertainty on the design to be a monotonous function over $\ve P$. We note here that the optimization for the worst case can lead to overly conservative results~\citep{lucia2013} and its practical benefits lie in the endeavors of seeking a feasible design with strict operational constraints of the experiments.

\subsection{Scenario-based Approach}
\label{subsec:scenario}
\begin{figure}
\centering
\psfrag{E1p}[tt][bb][0.8]{$\hat{\ve p}^1\ \ \ \ \ $}
\psfrag{E2p}[bb][cc][0.8]{$\ \ \ \ \hat{\ve p}^2$}
\psfrag{Esp}[cc][tt][0.8]{$\hat{\ve p}^{n_s}\ \ \ \ $}
\psfrag{u11}[cc][cc][0.8]{$\mathcal U^1_N$}
\psfrag{u21}[cc][cc][0.8]{$\mathcal U^2_N$}
\psfrag{us1}[cc][cc][0.8]{$\mathcal U^{n_s}_N$}
\psfrag{N}[cc][bb][0.8]{\!\!$N$}
\includegraphics[width=0.3\textwidth]{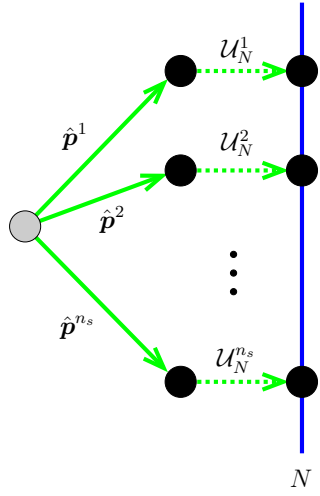}
\caption{Illustration of the scenario-based DoE.}
\label{fig:scenario}
\end{figure}

Opposed to the robust min-max design, the idea behind the scenario-based DoE~\citep{gal06, Telen_roed, wel09} is to optimize for the mean value of the objective under the stochastic realization of the uncertainty ($\ve p\in\ve P$). This approach considers $n_s$ discrete realizations (scenarios) of $\hat{\ve p}$ from the set $\ve P$ and is illustrated in Figure~\ref{fig:scenario}. The figure shows a scenario tree with:
\begin{itemize}
\item Grey node representing a priori estimation of the different expected values of parameters to be considered.
\item Solid arrows signifying the assignment of the different expected values of parameters to $n_s$ scenarios.
\item Black nodes and dashed arrows denoting the design of $N$ experiments, where the common color of the designs (green dashed arrows) expresses that the designs are mutual.
\end{itemize}

The optimization then seeks a common experiment policy for all the scenarios that minimizes the average---weighted by the scenario probability---of the objectives of the individual scenarios. The associated optimization problem is given as
\begin{subequations}\label{eq:scenario}
  \begin{align}
  \min_{\mathcal U_N^s, \forall s\in\{1,2,\dots,n_{s}\}}& \sum_{s=1}^{n_{s}} \omega^s \phi_A(\mathbb{FIM}(\hat{\ve p}^s, \mathcal U_N^s))\\
  \text{s.t. } & \forall \tau\in\{\tau_1,\tau_2,\ldots,\tau_N\}:\notag\\
  & \ve{u}^1_{\tau} = \ve{u}^2_{\tau} = 
  \cdots = \ve{u}^{n_s}_{\tau}, \label{eq:sc_nonanticipativity}\\
  &\forall \tau\in\{\tau_1,\tau_2,\dots,\tau_N\}, \ \forall s\in\{1,2, \ldots,n_{s}\}:\notag\\
  &\quad \hat{\ve y}(\hat{\ve p}^s, \tau)= \ve{F}(\hat{\ve p}^s, \ve{u}_{\tau}^s),\\
  &\quad \ve u^L\leq \ve u^s_\tau \leq \ve u^U,
  \end{align}
\end{subequations}
where $\omega^s$ represents the weight (probability with $\sum_{s=1}^{n_{s}} \omega^s=1$) of the $s^\text{th}$ scenario and $\hat{\ve p}^s$ is the particular realization of $\hat{\ve p}$ in the $s^\text{th}$ scenario.

In the simplest alternative, the scenarios can be selected as combinations of minimal, nominal and, maximal parameter values from $\ve P$ if one assumes a uniform probability distribution of $\hat{\ve p}$ over $\ve P$. If $\ve P$ represents a joint-confidence region, the scenarios can be selected to represent the (approximate) sigma points of $\ve P$~\citep{nim20}. This setup would give rise to the Bayesian DoE. One can also consider scenarios to be the samples from the underlying probability distribution within $\ve P$~\citep{mes15}, which can then be regarded as a stochastic DoE. In all the cases, one can assign a certain probability to the scenarios selected through the values of $\omega^s$.

We note that the scenario-based approach can also be a viable alternative to the min-max approach for DoE problems with hard constraints on system outputs or internal states such as studied in~\cite{pet20} and references therein. This is supported by the possibility of identifying the worst-case scenario---in terms of feasibility---that is often present in practical applications by one of the extreme points of $\ve P$ or can be pre-identified~\citep{hol19}. Such scenarios can then be added among the branches of the scenario tree to guarantee robust feasibility. Moreover, the optimization of the weighted objective of the scenarios---i.e., optimizing for the mean performance---would result in outperforming the min-max DoE by the scenario-based DoE on average.

\subsection{Multi-stage Approach}
\label{subsec:multistage}
The scenario-based and min-max DoE are one-shot (single-stage) approaches in principle---\ie all the $N$ experiments are conducted once the design is calculated. Of course, one can combine these approaches with the sequential approach for iterative re-design. The major issue within this approach is that the aforementioned designs do not consider explicitly the possibility of re-estimation and re-design (recourse) based on the available intermediate information.

\subsubsection{Two-stage Design of Experiments}
\begin{figure}
\centering
\psfrag{E1p}[tt][bb][0.8]{$\hat{\ve p}^1\ \ \ \ \ $}
\psfrag{E2p}[cc][tt][0.8]{$\ \ \ \ \hat{\ve p}^2$}
\psfrag{Esp}[cc][tt][0.8]{$\hat{\ve p}^{n_s}\ \ \ \ $}
\psfrag{u11}[cc][cc][0.8]{$\mathcal U^1_{N_e}$}
\psfrag{u21}[cc][cc][0.8]{$\mathcal U^2_{N_e}$}
\psfrag{us1}[cc][cc][0.8]{$\mathcal U^{n_s}_{N_e}$}
\psfrag{E1p1}[cc][cc][0.8]{$\hat{\ve p}^1$}
\psfrag{E2p1}[cc][cc][0.8]{$\hat{\ve p}^2$}
\psfrag{Esp1}[cc][cc][0.8]{$\ \hat{\ve p}^{n_s}$}
\psfrag{u1np}[c][c][0.8]{$\!\mathcal U^{1}_{N-N_e}$}
\psfrag{u2np}[c][c][0.8]{$\!\mathcal U^{2}_{N-N_e}$}
\psfrag{usnp}[c][t][0.8]{$\mathcal U^{n_s}_{N-N_e}$}
\psfrag{Nms}[cc][bb][0.8]{\!\!\!\!$N_e$}
\psfrag{N}[cc][bb][0.8]{\!\!$N$}
\includegraphics[width=0.6\textwidth]{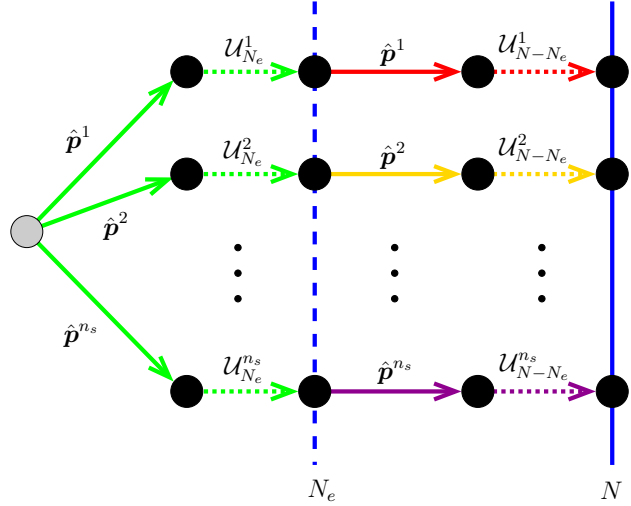}
\caption{Illustration of the two-stage DoE.}
\label{fig:two-stage}
\end{figure}

The explicit consideration of the so-called recourse actions~\citep{Garstka1974} gives rise to multi-stage decision making. Such a scheme effectively combines elements of the sequential and scenario-based DoE. We will build the idea of the multi-stage DoE incrementally---starting with the simplest variant---i.e., the two-stage DoE. We illustrate the approach through a decision (scenario) tree in Figure~\ref{fig:two-stage}, which can be easily compared with Figure~\ref{fig:scenario} to observe the differences between the approaches. As in the scenario-based DoE, we explicitly model the effect of uncertainty by considering $n^s$ designs (scenarios) each with a different realization of $\hat{\ve p}$. Unlike in the scenario-based DoE, we only require that the first $N_e$ experiments are common for all the scenarios. This models the so-called non-anticipativity of the design with only a priori information and one can think of this as the sequential design considering the uncertainty in the a priori information. After designing $N_e$ experiments---i.e., the first-stage decision (marked by the blue dashed line in the figure)---each design (branch) can be freely optimized (second-stage decision) for the respective value of $\hat{\ve p}:=\hat{\ve p}^k, k\in\{1, 2, \dots, n_s\}$, which models the re-estimation procedure (as done in the implementation of the sequential design) that reveals one of the scenarios as being true. The decoupling of the scenarios is illustrated by different colors of the branches after the stage $N_e$.

The two-stage DoE takes into account both the presented stages and finds the best conditions for initial $N_e$ experiments, which make the successive $N-N_e$ re-designed (recourse) experiments optimize the weighted-average performance of the outcome from all $N$ experiments. The underlying optimization problem for the two-stage DoE reads as
\begin{subequations}\label{eq:2st}
  \begin{align}
  \min_{\mathcal U_N^s, \forall s\in\{1,2,\dots,n_{s}\}}& \sum_{s=1}^{n_{s}} \omega^s \phi_A(\mathbb{FIM}(\hat{\ve p}^s, \ve u^s))\\
  \text{s.t. } & \forall \tau\in\{\tau_1,\tau_2,\ldots,\tau_{N_e}\}:\notag\\
  &\ve{u}^1_{\tau} = \ve{u}^2_{\tau} = 
  \cdots = \ve{u}^{n_s}_{\tau}, \label{eq:2st_nonanticipativity}\\
  &\forall \tau\in\{\tau_1,\tau_2,\dots,\tau_N\}, \ \forall s\in\{1,2,\ldots,n_{s}\}:\notag\\
  &\hat{\ve y}(\hat{\ve p}^s, \tau)= \ve{F}(\hat{\ve p}^s, \ve{u}_{\tau}^s),\\
  &\ve u^L\leq \ve u^s_\tau \leq \ve u^U.
  \end{align}
\end{subequations}
Note that the mathematical formulation represents only a minor modification of the scenario-based approach~\eqref{eq:scenario} regarding the so-called non-anticipativity constraints (compare~\eqref{eq:sc_nonanticipativity} and~\eqref{eq:2st_nonanticipativity}).

This formulation looks at identifying the best possible experiments in the early phase of experimentation, when the parameter knowledge is poor and vice versa. In our previous work, we suggested a similar strategy in the context of set-membership (guaranteed) estimation~\citep{anwesh17esc}, which offers an even more intuitive explanation of the effects of multi-stage decision making on experiment design.

The allocation of experiments to the first and second stage is an adjustable parameter. If no prior experimental results are available---\ie one only possesses the knowledge of $\ve P$---he/she would naturally set $N_e$ in the first stage such that $N_e\times n_y\geq n_p$ to satisfy the standard minimal identifiability conditions. We note that there are other criteria for selecting the least number of experiments~\citep{geo13}, yet a study of these aspects is beyond our scope in this paper. A simulation-based tuning is also possible here in combination with some previous plant expertise. The case, in which past experiments are available, naturally opens up wider options to tune $N_e$ especially in situations, where $N$ is relatively small.

The application of the two-stage design can either be done in an open-loop or a closed-loop manner. In the open-loop approach, re-estimation is performed at the stage $N_e$---\ie after the experiments designed in the first stage are performed. This is followed by a (nominal) re-design and realization of the remaining experiments. In the closed-loop approach the two-stage DoE is applied in a moving-horizon fashion---\ie it is recast from the stage $N_e$ repetitively. The decision of performing an open- or closed-loop approach is mainly connected to the number of available measurements $N$ and the quality of estimates obtained at the stage $N_e$---similar to the workflow of the sequential DoE.

The application of the two-stage DoE is outlined in the following pseudo-algorithm:
\begin{enumerate}
 \item For the given model, $n_p$, and $n_y$, choose $N$ (if not given), $N_e$ (if not given), and $n_s$.
 \item For the given $\ve P$, choose $\hat{\ve p}^s\in\ve P, \forall s\in\{1, 2, \dots, n_s\}$ and solve the problem~\eqref{eq:2st} to design $N$ experiments.
 \label{it:alg_2st_DoE}
 \item Apply the resulting $\mathcal U_{N_e}^*$ and collect the measurements $\ve y_m$.
 \item Identify new $\hat{\ve p}$ and $\ve P$ using the least-squares estimation (Section~\ref{sec:lse}) over all the conducted experiments.
 \item Update $N:=N-N_e$.
 \item Go to step~\ref{it:alg_2st_DoE} unless:
 \begin{enumerate}
 \item implementation is done in open-loop fashion, or
 \item $N \leq N_e$, or
 \item only a marginal (user-defined) improvement of the confidence region---measured by the chosen design criterion---was reached, or
 \item only a marginal (user-defined) update of the least-squares estimates was obtained.
 \end{enumerate}
 \item Use the nominal DoE to design the remaining $N$ experiments for the current value $\hat{\ve p}$. Terminate.
\end{enumerate}
The proposed algorithm can be regarded as a combination of the sequential and scenario-based (stochastic or Bayesian) DoE. Recall that the scenario-based approach can also effectively replace the min-max approach for a suitable choice of the scenarios. Thus the multi-stage approach combines the above designs. Moreover, multi-stage decision making enables the consideration of recourse actions by future experiments. The price to pay for this asset lies in the increased problem complexity---i.e., an increased number of optimized variables. We note that the use of the nominal DoE as a final step is not mandatory and it can be replaced by scenario-based DoE or min-max DoE.

A situation when the experiment-design problem at hand features strict operational constraints is not in our scope in this study. However, as mentioned earlier regarding (approximate) min-max and scenario-based approaches, the feasibility of the design regarding the operational constraints can be incorporated into the consideration straightforwardly.

The two-stage DoE is more optimistic than the scenario-based DoE as its formulation assumes that the true realization of the uncertainty is revealed after $N_e$ experiments. This on one hand means that the worst-case performance of the multi-stage DoE can be expected to be worse than the performance of the scenario-based approach. On the other hand, if parameter re-estimation (and experiment re-design) is possible in-between the stages, the multi-stage DoE will outperform the scenario-based DoE.

The optimistic nature of the two-stage DoE can be made more realistic by two possible modifications. The first option lies in the usage of techniques from a so-called dual control, where the essential part of the technology is the stochastic prediction of future estimates~\citep{filatov04}. As an example, \cite{Thangavel2018c} illustrated a dual-control approach in the framework of multi-stage NMPC, which is directly applicable here. The second option represents the use of the multi-stage DoE, where one continues to branch the tree after $N_e$ experiments and the optimization thus accounts for the uncertainty in the recourse experiments. 

\subsubsection{Multi-stage Design of Experiments}
As mentioned above, by increasing the number of stages beyond two---i.e., by continuing to branch the tree after the first stage---one can model the uncertainty in the parameter estimates after processing the first $N_e$ experiments. The multi-stage DoE is formulated over $n_r+1$ stages, where the tree branches in the first $n_r$ (robust) stages. Similarly to the two-stage approach, one has to decide about the number of performed experiments in each stage $i$. Here, $N_e^i$ denotes the total number of experiments conducted until the stage $i$ and it holds that $N_e^1 < N_e^2 < \dots < N_e^{n_r} < N_e^{n_r+1}\equiv N$.

\begin{figure}
\centering
\psfrag{01}[cc][cc][0.8]{$\hat{\ve p}^{1,1}$}
\psfrag{2}[cc][tt][0.8]{$\mathcal U_{N_e^1}^{1,1}$}
\psfrag{3}[cc][cc][0.8]{$\hat{\ve p}^{1,2}$}
\psfrag{4}[cc][tt][0.8]{$\mathcal U_{N_e^1}^{1,2}$}
\psfrag{5}[cc][cc][0.8]{$\hat{\ve p}^{2,1}$}
\psfrag{6}[cc][tt][0.8]{\!$\mathcal U_{N_e^2-N_e^1}^{2,1}$}
\psfrag{7}[cc][cc][0.8]{$\hat{\ve p}^{2,2}$}
\psfrag{8}[cc][tt][0.8]{\!$\mathcal U_{N_e^2-N_e^1}^{2,2}$}
\psfrag{9}[cc][cc][0.8]{$\hat{\ve p}^{2,3}$}
\psfrag{10}[cc][tt][0.8]{\!\!\!$\mathcal U_{N_e^2-N_e^1}^{2,3}$}
\psfrag{1}[cc][cc][0.8]{$\hat{\ve p}^{2,4}$}
\psfrag{12}[cc][tt][0.8]{\!\!\!$\mathcal U_{N_e^2-N_e^1}^{2,4}$}
\psfrag{13}[cc][cc][0.8]{$\hat{\ve p}^{3,1}$}
\psfrag{14}[cc][tt][0.8]{\!$\mathcal U_{N-N_e^2}^{3,1}$}
\psfrag{15}[cc][cc][0.8]{$\hat{\ve p}^{3,2}$}
\psfrag{16}[cc][tt][0.8]{\!$\mathcal U_{N-N_e^2}^{3,2}$}
\psfrag{17}[cc][cc][0.8]{$\hat{\ve p}^{3,3}$}
\psfrag{18}[cc][tt][0.8]{\!$\mathcal U_{N-N_e^2}^{3,3}$}
\psfrag{19}[cc][cc][0.8]{$\hat{\ve p}^{3,4}$}
\psfrag{20}[cc][tt][0.8]{\!$\mathcal U_{N-N_e^2}^{3,4}$}
\psfrag{34}[cc][cc][0.8]{$N_e^1$}
\psfrag{21}[cc][cc][0.8]{$N_e^2$}
\psfrag{22}[cc][cc][0.8]{$N$}
\psfrag{23}[cc][cc][0.8]{}
\psfrag{24}[cc][cc][0.8]{}
\psfrag{25}[cc][cc][0.8]{}
\psfrag{26}[cc][cc][0.8]{}
\psfrag{27}[cc][cc][0.8]{}
\psfrag{28}[cc][cc][0.8]{}
\psfrag{29}[cc][cc][0.8]{}
\psfrag{30}[cc][cc][0.8]{}
\psfrag{31}[cc][cc][0.8]{}
\psfrag{32}[cc][cc][0.8]{}
% \psfrag{23}[cc][cc][0.8]{$\ve \omega^{1,1}$}
% \psfrag{24}[cc][cc][0.8]{$\ve \omega^{1,2}$}
% \psfrag{25}[cc][cc][0.8]{$\ve \omega^{2,1}$}
% \psfrag{26}[cc][cc][0.8]{$\ve \omega^{2,2}$}
% \psfrag{27}[cc][cc][0.8]{$\ve \omega^{2,3}$}
% \psfrag{28}[cc][cc][0.8]{$\ve \omega^{2,4}$}
% \psfrag{29}[cc][cc][0.8]{$\ve \omega^{3,1}$}
% \psfrag{30}[cc][cc][0.8]{$\ve \omega^{3,2}$}
% \psfrag{31}[cc][cc][0.8]{$\ve \omega^{3,3}$}
% \psfrag{32}[cc][cc][0.8]{$\ve \omega^{3,4}$}
\psfrag{33}[cc][cc][0.8]{$\hat{\ve p}$}
\includegraphics[width=\textwidth]{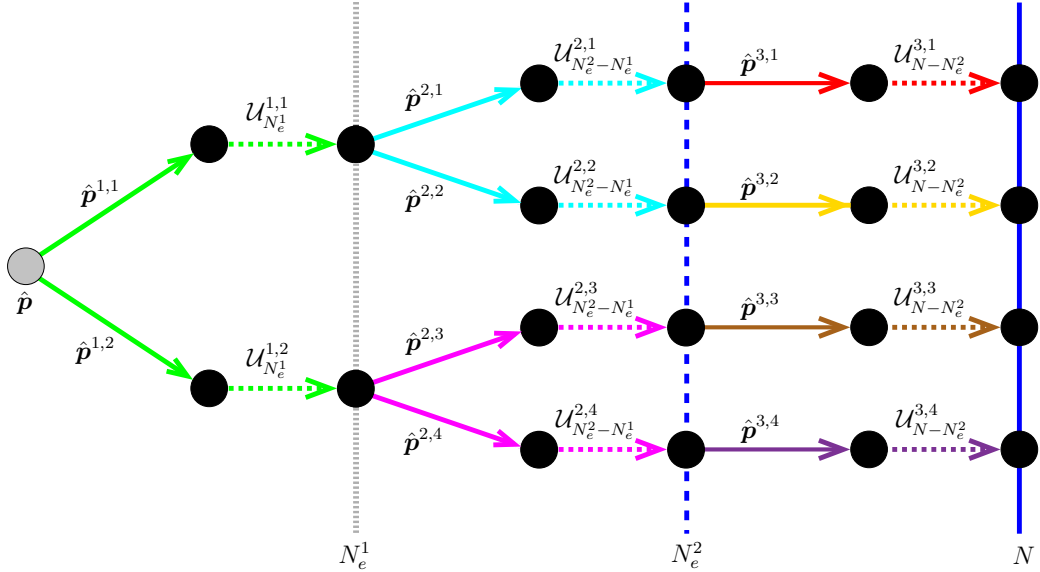}
\caption{Illustration of the multi-stage DoE.}
\label{fig:multi-stage}
\end{figure}

We illustrate the difference between the two-stage and multi-stage DoE on an example of three-stage DoE ($n_r=1$) in Figure~\ref{fig:multi-stage}, which uses the same visual coding as Figure~\ref{fig:two-stage}. In each of the first two stages, the incoming (estimated) $\hat{\ve p}$ is discretized into $n_s := 2$ branches. After $n_r$ stages (denoted by $N_e^{n_r}\equiv N_e^2$), a sequential-design principle is applied over the individual scenarios. This is represented by the solid parallel branches of different colors. This illustrates that the decisions are decoupled and that $n_s^{n_r}$ individual re-designs are performed. The number of experiments conducted until the $i^\textsuperscript{th}$-stage ($N_e^i$) is a degree of freedom to be chosen. The overall optimization takes all stages into account and finds the best conditions for experiments in each stage, which make the successive re-designed (recourse) experiments to minimize the weighted-average performance of the outcome from $N$ experiments. 

The underlying optimization problem is formulated as
\begin{subequations}\label{eq:ms_new}
  \begin{align}
  \min_{\stackrel{\mathcal U_{N_e^i-N_e^{(i-1)}}^{i,s},\ \mathcal U_{N-N_e^{n_r}}^{n_r+1,s}}{\forall i \in \{1,2,\ldots, n_r\}, \forall s \in \{1,2,\ldots,n_s^i\}}}& \sum_{i=1}^{n_r} \sum_{s=1}^{n_s^i} 
  \omega^{i,s} \phi_A\left(\mathbb{FIM}\left(\hat{\ve p}^{i,s}, \mathcal U_{N_e^i-N_e^{(i-1)}}^{i,s}\right)\right)\notag\\
  &\qquad+\sum_{s=1}^{n_s^{n_r}} \omega^{n_r+1,s} \phi_A\left(\mathbb{FIM}\left(\hat{\ve p}^{n_r+1,s}, \mathcal U_{N-N_e^{n_r}}^{n_r+1,s}\right)\right)\\
  \text{s.t. } & \forall i \in \{1,2,\ldots, n_r\}, \forall s \in \{1,2,\ldots,n_s^i\}, \notag\\ 
  &\forall \tau \in \{\tau_{1+\sum_{k=1}^{i-1} N_e^k}, \tau_{2+\sum_{k=1}^{i-1} N_e^k}, \ldots, \tau_{\sum_{k=1}^{i} N_e^k}\}:\notag\\
  &\quad \ve u_{\tau}^{i,s} = \ve u_{\tau}^{i,\hat s}, \ \text{if } \text p(s)= \text p(\hat s), \label{eq:ms_nonanticipativity_new}\\
  &\quad \hat{\ve y}(\hat{\ve p}^{i,s}, \tau)= \ve{F}(\hat{\ve p}^{i,s}, \ve{u}_{\tau}^{i,s}),\\
  &\quad \ve u^L\leq \ve u^{i,s}_\tau \leq \ve u^U,\\
  & \forall s \in \{1,2,\ldots,n_s^{n_r}\}, \forall \tau \in \{\tau_{N_{e}^{n_r}+1}, \tau_{N_{e}^{n_r}+2}, \ldots, \tau_{N}\}:\notag\\
  &\quad \hat{\ve y}(\hat{\ve p}^{n_r+1,s}, \tau)= \ve{F}(\hat{\ve p}^{n_r+1,s}, \ve{u}_{\tau}^{n_r+1,s}),\\
  &\quad \ve u^L\leq \ve u^{n_r+1,s}_\tau \leq \ve u^U,
  \end{align}
\end{subequations}
where $\text p(s)$ denotes the parent node of the node $s$ and where $N_e^{0}$ is interpreted as 0. The multi-stage DoE enjoys practically the same characteristics as the two-stage DoE. It represents a better model of the decision when closed-loop implementation of the design is performed and is more precise.

The weights $\omega^{i,s}$ can be chosen in the same fashion as in the scenario-based and two-stage designs, i.e., for each $i$, $\sum_{s=1}^{n_s^i}\omega^{i,s}=1$. In principle, the weights should also respect that for each $i\leq n_r$, $\omega^{i, s} = \sum_{\forall \hat s, \hat s=\text p(s)}\omega^{i+1,\hat s}$. This is the simplest consistent choice if no a priori information is known regarding the propagation of the underlying (assumed) probability distribution through the nonlinear system.

\section{Case Studies}\label{sec:case}
We test the presented methodologies on four nonlinear parameter estimation problems, three simple ones, where we analyze the performance of the presented approaches, and one real-world problem. As the scenario-based designs consider discretization of the set $\ve P$ into a finite number of scenarios, we implement the same strategy in the min-max approach to make a fair comparison among these approaches. From the multi-stage DoE approaches, we only apply the two-stage in an open-loop fashion to illustrate the main benefits of the proposed approach. The use of two-stage DoE is relevant because of the relative simplicity of the case studies regarding the number of available experiments.

The presented rDoE methodologies are evaluated in the following setup:
\begin{itemize}
 \item Nominal, min-max, and scenario-based approaches
 
 As these approaches use no re-estimation and re-design, the experiments designed according to the $\ve P$ are directly applied.
 \item Sequential and two-stage approaches
 \begin{enumerate}
 \item In the first stage, the parameters are only known to lie in the set $\ve P$. A designed experiment is thus executed until $N_e$ measurements are obtained.
 \item After taking $N_e$ measurements, it is assumed that the least-squares estimation reveals true values of the parameters.
 
 This serves for a fair comparison of the designs as it mitigates the effects connected to the actual noise realization. One can thus interpret the results as comparing the mean performance---w.r.t. the measurement noise---of the robustification methods.
 \end{enumerate} 
\end{itemize}
The value of optimal design criterion---i.e., a "crystal-ball" solution $\phi_A^* = \phi_A(\mathbb{FIM}(\ve p^*, \ve u^*))$---serves as a reference and is compared to the values reached by applying the calculated design $\ve u$ for each experiment $\phi_A = \phi_A(\mathbb{FIM}(\ve p^*, \ve u))$.

The computational aspects of the different designs are elaborated in the discussion section.

\subsection{Case Study 1}\label{sec:cs_1par}
We consider one of the simplest nonlinear estimation problems, where one identifies only one parameter. The mathematical model of the system is stated as
\begin{align}\label{eq:casestudy_BOD_1par}
 \hat y(p, \tau) = 1-\exp(-p u_\tau).
\end{align}
This estimation can be regarded as an identification of a time constant $1/p$ of a first-order linear time-invariant dynamic system with a unitary static gain from its step response. Parameter $p$ can also be interpreted as a rate of decay.

The true parameter value is unknown but is assumed to lie within $p^*\in\ve P:=1+[-\Delta, \Delta]$. The nominal value of the parameter $\hat p$ is taken as the midpoint of $\ve P$. We consider the measurement error to be a random variable distributed as zero-mean, white Gaussian noise with standard deviation $\sigma = 0.1/3$. We use the simplest possible setup regarding the number of consecutive experiments and the number of experiments considered. This means that two consecutive experiments are to be conducted with one measurement obtained in each experiment---i.e., $N:=2$ and $N_e:=1$.

The optimal experiment design is very simple---i.e., $u^*=1/p^*$ for any number of repeated experiments. This shows the heavy dependency of DoE on the parametric values that are considered as expected. The different design approaches are calculated by implementing the corresponding optimization problems using the solver BARON~\citep{baron} and its Matlab interface with default options (e.g., relative optimality gap is set to $10^{-9}$). Min-max, scenario-based and two-stage approaches use three scenarios with a minimal, maximal, and nominal value of the parameter given by $P$. The performance of the different designs is evaluated using Monte Carlo simulations, where we vary the true value of parameters for each simulated design 100 times. The parameter values are taken from a uniformly sampled $\ve P$.

\begin{figure}
\centering
\psfrag{delta}[cc][cc][0.8]{$p^*-\hat p$}
\psfrag{loss}[cc][cc][0.8]{$\phi_A-\phi_A^*$}
\includegraphics[width=0.8\linewidth]{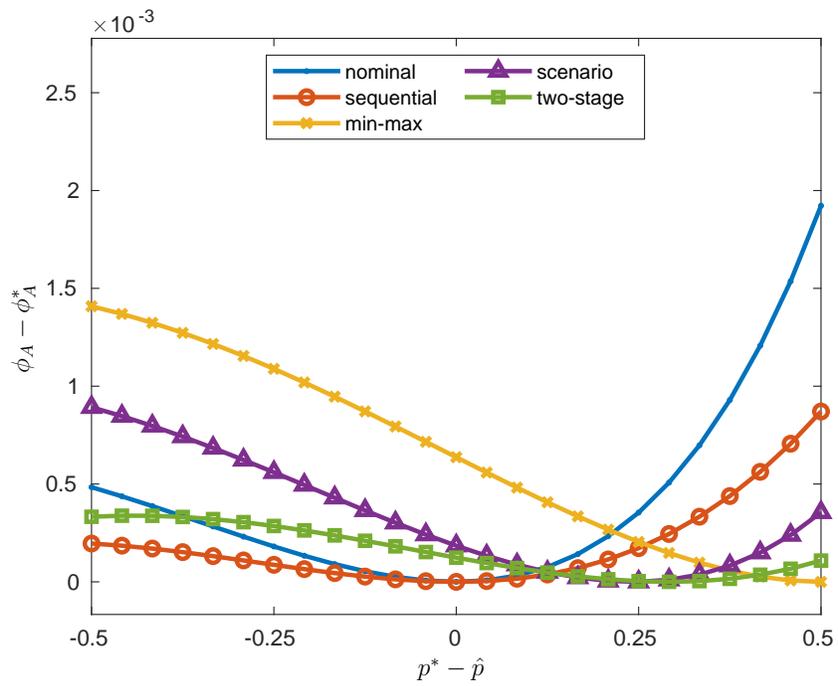}
\caption{Comparison of different DoE methods for Case study~1 with $\Delta=\frac{1}{2}$.}
\label{fig:ex1_res_del05}
\end{figure}

Figure~\ref{fig:ex1_res_del05} shows results in terms of optimality loss of the different designs ($\phi_A-\phi_A^*$) {vs.} the deviation of the true parameter value $p^*$ from its nominal value $\hat p$ obtained for the case when $\Delta=\frac{1}{2}$. It is apparent that different designs show the best performance values in different situations as expected. The nominal and sequential designs show the best performance if the true value of $p$ is close to its nominal value.

Min-max design optimizes for the worst-case and is naturally superior in this situation, which occurs for the largest values of $p^*$---i.e., at $p^*-\hat p=\frac{1}{2}$. The worst case appears at this point as a consequence of the confidence region size increase with increasing $p^*$. This is natural and expected as it is harder to detect the faster rate of decay ($p^*$) from the measurements compared to the case of slow decay---i.e., at $p^*-\hat p<0$.

The sequential and two-stage designs optimize the weighted average of the costs of the aforementioned scenarios (nominal and worst-case) together with the scenario of the smallest value of $p^*$ (the slowest decay)---i.e., at $p^*-\hat p=-\frac{1}{2}$. The design cost increases with increasing $p^*$, which occurs due to the aforementioned confidence region size increase. This causes the resulting designs to be truly optimal for the above-the-mean values of $p^*-\hat p$ (around $p^*-\hat p=\frac{1}{4}$). Note that one would be able to make the optimality loss more uniform here by using the insight obtained from this analysis and by tuning the weights of the different scenarios correspondingly. This shows the promising flexibility of the advanced design methods.

By observing Figure~\ref{fig:ex1_res_del05}, we can further confirm the results of the earlier theoretical analyses:
\begin{itemize}
 \item Worst-case performance of the sequential DoE is upper-bounded by the performance of the nominal DoE.
 \item The min-max DoE is an overly conservative robustification technique.
 \item The two-stage DoE improves the scenario-based DoE on average.
\end{itemize}

The presented plot further reveals the worst-case performance. As can be expected the nominal DoE scores the worst here. Its worst-case (and also overall) performance is not much different from the min-max DoE. While sequential and scenario-based approaches outperform nominal and min-max techniques---one should bear in mind here that the sequential approach can take decent advantage from re-estimation/re-design knowing the true value of $p$---they do not perform very well overall. Finally, we observe that the two-stage approach outperforms the min-max and scenario-based DoE and that there are many cases where it performs better than nominal and sequential designs. Also, this approach is evaluated optimistically (re-estimation/re-design is possible using the true value of $p$). If no re-estimation is possible, the two-stage DoE performs almost identically to the scenario-based approach, which, restores $53\,\%$ of optimality in the worst case and $17\,\%$ of optimality on average compared to the nominal DoE.

The average performance of the presented approaches can be summarized in percentages of relative suboptimality as $100\,\%$ (nominal DoE), $46\,\%$ (sequential DoE), $159\,\%$ (min-max DoE), $81\,\%$ (scenario-based DoE), and $38\,\%$ (two-stage DoE). This further documents the observed tendencies and clearly shows that the two-stage approach is superior compared to other designs. The performance of scenario-based and two-stage approaches can be further improved by increasing the number of scenarios considered. If the number of scenarios is increased from three to five---if we include the scenarios for $p^*-\hat p=-\frac{1}{4}$ and $p^*-\hat p=\frac{1}{4}$\,---the mean and the worst-case performance of these approaches improves by around 10\%.

\begin{figure}
\centering
\psfrag{delta}[cc][cc][0.8]{$p^*-\hat p$}
\psfrag{loss}[cc][cc][0.8]{$\phi_A-\phi_A^*$}
\includegraphics[width=0.8\linewidth]{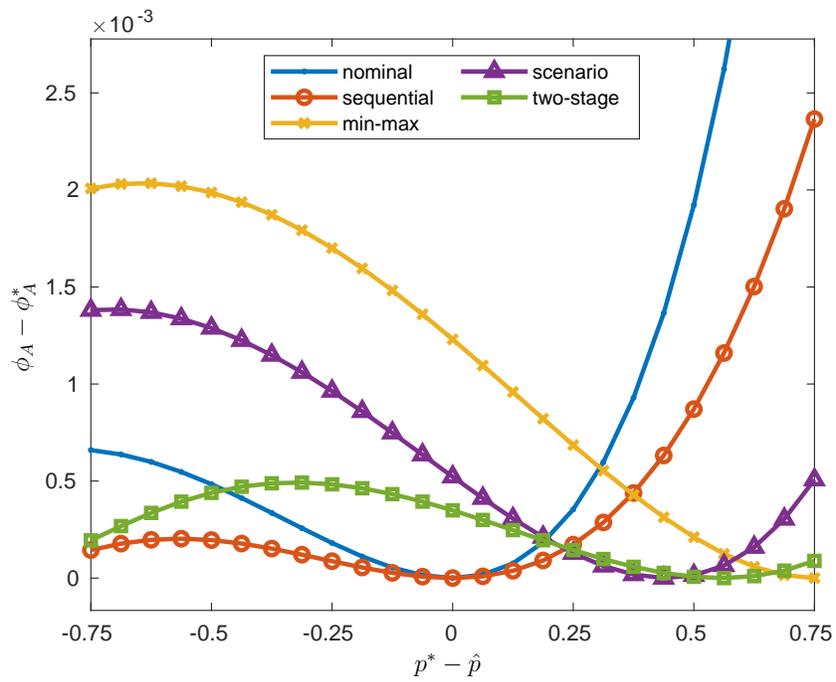}
\caption{Comparison of different DoE methods for Case study~1 with $\Delta=\frac{3}{4}$.}
\label{fig:ex1_res_del075}
\end{figure}

Next, we study a scenario of an increased level of uncertainty, where $\Delta$ is increased from $\frac{1}{2}$ to $\frac{3}{4}$. The results are visualized in Fig.~\ref{fig:ex1_res_del075}. As can be expected, the mean as well as the worst-case performance of all approaches deteriorates. Compared to the case of $\Delta=\frac{1}{2}$, the drop in optimality loss is most significant for the nominal DoE (mean: $155\,\%$, worst-case $203\,\%$) and for the sequential DoE (mean: $132\,\%$, worst-case $172\,\%$). The scenario-driven approaches exhibit much less sensitivity to the increased level of uncertainty. Their performance worsens for mean and worst-case performance, respectively, by $73\,\%$ and $45\,\%$ (min-max DoE), $93\,\%$ and $55\,\%$ (scenario-based DoE), and by $64\,\%$ and $46\,\%$ (two-stage DoE).

Let us also shed light on the quality of the resulting parametric confidence regions (intervals in this case). The $2\sigma$-confidence intervals would bear an uncertainty of $\pm0.06$ and of $\pm0.08$ if the experiments are designed using the two-stage and sequential approaches, respectively, if the worst-case is realized. While this represents an improvement of 25\%, the absolute improvement is not much significant. We note here, that in the case of subsequent estimation with unknown variance, the confidence intervals enlarge to roughly $\pm0.4$ and $\pm0.6$, respectively, which is already considerable.

% CPU time
% nom/seq mmax sc ms 
% 0.0850 0.1304 2.8960 2.0444

% chi2inv(0.95, 1)
% 3.8415
% 
% finv(0.95, 1, 1)
% 161.4476

% % if 3 scenarios
%          0    0.3705    0.2100    1.7301
%          0    0.1710    0.1030    0.7835
%     0.0000    0.5906    0.5733    1.2676
%     0.0002    0.3000    0.2173    0.8031
%     0.0000    0.1407    0.1113    0.3042

% % if uncertainty increased by 50%
%          0    0.9439    0.4355    5.2432
%          0    0.3963    0.1597    2.1284
%     0.0000    1.0193    1.1070    1.8310
%     0.0001    0.5799    0.4699    1.2458
%     0.0000    0.2310    0.2410    0.4429

% % if 5 scenarios
%          0    0.3705    0.2100    1.7301
%          0    0.1710    0.1030    0.7835
%     0.0000    0.5906    0.5733    1.2676
%     0.0007    0.2785    0.2313    0.7248
%     0.0005    0.1256    0.1190    0.2741

\subsection{Case Study 2}\label{sec:cs_stat}
% CPU time
% linsys 1 ord
% 
% cputim_nom =
% 
%     2.7314
% 
% cputim_minmax =
% 
% 0.5513
% 
% cputim_sc =
% 
%   162.0713
% 
% cputim_ms =
% 
%    15.0870
% 
% 
% reacABC
% 
% cputim_nom =
% 
%     2.6818
% 
% cputim_mmax =
% 
%     1.1487
% 
% cputim_sc =
% 
%   104.7113
% 
% cputim_ms =
% 
%    35.5861
% 
% 
% temeprature growth
% cputim_nom =
% 0.0363
% 
% cputim_mmax =
% 1.9854
% 
% cputim_sc = 
% 1.3803
% 
% cputim_ms =
% 1.6375e+03

We consider an example that is an extension of Case study~1, where we include estimation of one more parameter that enters the model linearly. 
The output of the studied system can be modeled by:
\begin{align}\label{eq:casestudy_BOD}
\hat y(\ve p, \tau) = p_1(1-\exp(-p_2 u_\tau)), \quad u_\tau \in [0,20],
\end{align}
where the parameter $p_1$ can be interpreted as a static gain of a first-order linear time-varying system with a time constant $1/p_2$. The model would thus describe a (unit) step response of the dynamic system. The example allows also an interpretation of the classical mathematical model for biological oxygen demand (BOD)~\citep{bat88}. As our further analysis will reveal, the bounds on $u$ are only formal as a) any reasonable experiment would take positive values and b) measurements taken beyond $u\geq10$ are practically identical as---after a period of ten time constants---the steady state is practically reached.

The true parameter values are unknown but are assumed to lie within $\ve P:=[0.5, 1.5]\times[0.5, 1.5]$. Note that we consider the same level of uncertainty as studied (in the first sub-case) in Case study~1 regarding the parameter $p_2$. The nominal values of parameters are taken as the midpoint of $\ve P$. We consider the same measurement error as in the previous case study. We use again the simplest setup regarding the number of available experiments, taking two measurements in two consecutive experiments---i.e., $N:=4$ and $N_e:=2$.

\begin{figure}
\centering
\psfrag{p1}[cc][cc][0.8]{$p_1$}
\psfrag{p2}[cc][cc][0.8]{$p_2$}
\includegraphics[width=0.8\linewidth]{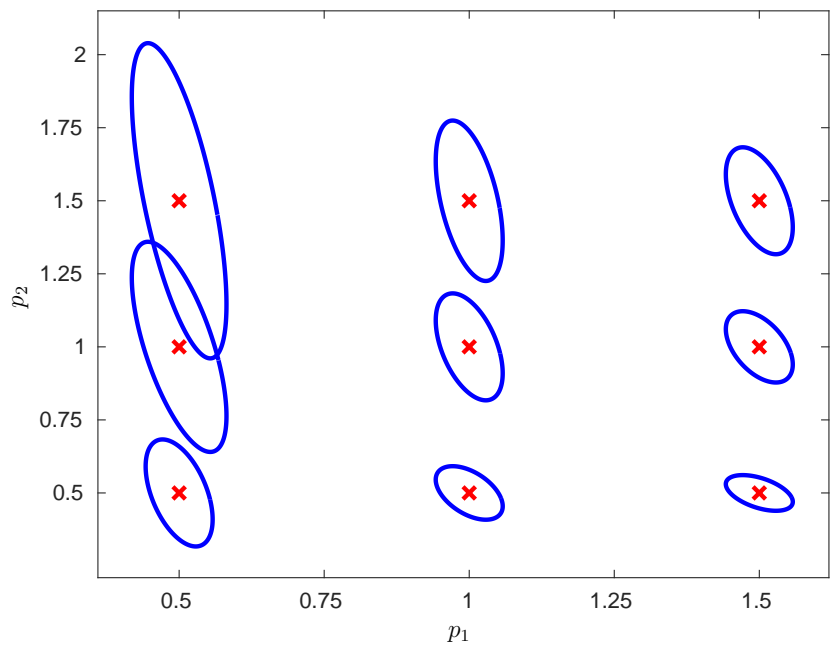}
\caption{Comparison of 2$\sigma$-joint-confidence regions for different values of the parameters for Case study~2.}
\label{fig:conf_reg_comp_1stord}
\end{figure}

To demonstrate the nature of the case study, we calculate the nominal designs for $N:=4$ corresponding to all combinations of minimal, nominal, and maximal values of parameters within $\ve P$. The general structure of the designs is to repeat a two-measurement experiment with $u_\tau=20$, which mainly helps to estimate $p_1$ (static gain), and $u_\tau\in[0.45, 1.70]$, which primarily regards into estimation of $p_2$ (reciprocal time constant). It can be concluded that the parametric uncertainty prominently affects the estimation of $p_2$, which is reminiscent of the previous case study. For further analysis, we plot the optimal confidence regions that correspond to the optimal designs in Figure~\ref{fig:conf_reg_comp_1stord}. We can observe that the nonlinearity of the estimation problem in $p_2$ causes large variance along this dimension in the size of the confidence region. At this point, we note that there are certain differences in the designs for different values of $N$. In other words, the values of $\ve u^*$ depend also on the number of designed experiments. This fact supports the use of designs aware of the number of possible experiments, unlike the myopic designs such as the sequential DoE.

\begin{figure}
\centering
\psfrag{delta}[cc][tt][0.8]{$\phi_A^{\text{seq}}-\phi_A^{\text{2-stg}}$}
\psfrag{p1}[cc][cc][0.8]{$p_1$}
\psfrag{p2}[cc][cc][0.8]{$p_2$}
\psfrag{SEQUENTIDOMINATES}[cc][cc][0.8]{$\!\!\phi_A^{\text{seq}}-\phi_A^{\text{2-stg}} < 0$}
\psfrag{TWOSTAGEDOMINATES}[cc][cc][0.8]{$\!\!\phi_A^{\text{seq}}-\phi_A^{\text{2-stg}} > 0$}
\includegraphics[width=0.8\linewidth]{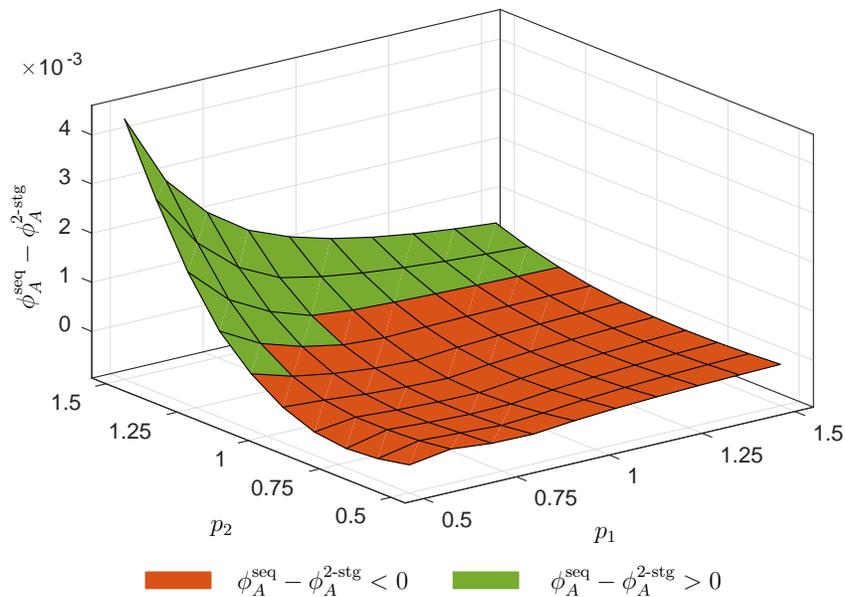}
\caption{Optimality-loss comparison of the sequential ($\phi_A^{\text{seq}}$) and the two-stage ($\phi_A^{\text{2-stg}})$ approaches to robust A design for different true values of parameters for Case study~2.}
\label{fig:perf_seq_2stg_comp_Ades_1stord}
\end{figure}

The computational setup---regarding the use of BARON and tuning and evaluation of rDoE approaches---is the same as in Case study~1. We first evaluate the performance of the two most promising approaches from the previous case study, the sequential and two-stage DoE. Figure~\ref{fig:perf_seq_2stg_comp_Ades_1stord} shows the corresponding results from Monte Carlo simulation with varied true values of the parameters. For clarity, the plot is constructed such that it shows the benefit of the two strategies compared to each other, which is expressed in terms of difference of the objective values reached in reality---i.e., $\phi_A^{\text{seq}}-\phi_A^{\text{2-stg}}$. In the plot, negative values (marked in red) reveal cases where it is better to use the sequential DoE than the two-stage design. This holds vice versa for the positive values (marked in green). The conclusions implied by the plot are similar to those observed for the one-dimensional problem (see Case study~1). The sequential DoE is the best in the median performance and in the best case---i.e., when the true values of the parameters are close to the nominal ones. Also, the cases, where the performance of the sequential DoE is superior, are related to the instances of true parameter values, where the estimation is more effective---i.e., where the small value of $p_2$ results in small confidence regions.

\begin{figure}
\centering
\psfrag{delta}[cc][tt][0.8]{$\phi_A-\phi_A^*$}
\psfrag{strat}[tt][bb][0.8]{$\leftarrow$ DoE strategy $\rightarrow$}
\psfrag{nom}[cc][bb][0.8]{nominal}
\psfrag{seq}[cc][bb][0.8]{sequential}
\psfrag{mmax}[cc][bb][0.8]{min-max}
\psfrag{rob}[cc][bb][0.8]{scenario}
\psfrag{rob2}[cc][bb][0.8]{two-stage}
\includegraphics[width=0.8\linewidth]{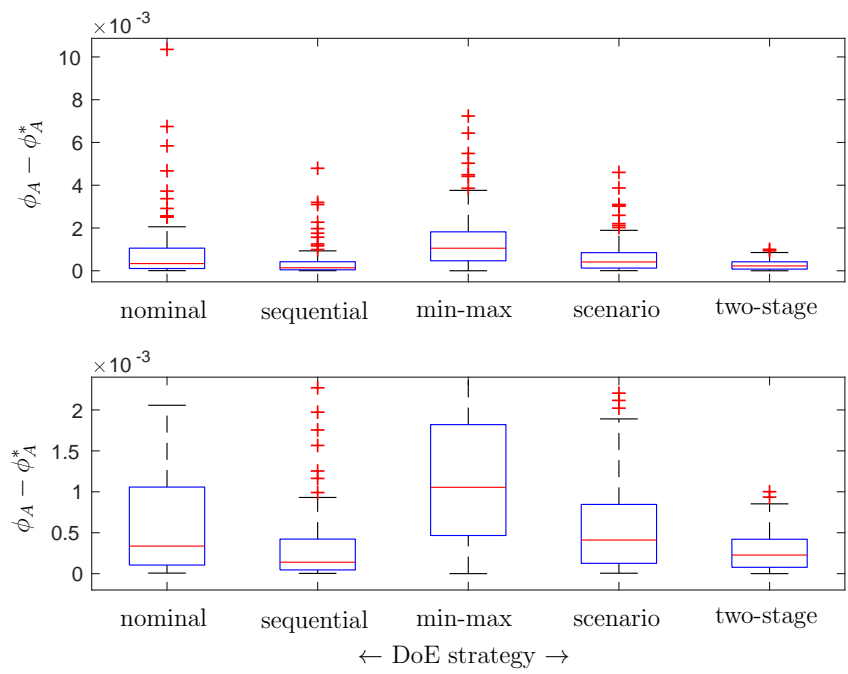}
\caption{A box plot (top -- full size, bottom -- zoomed in) of optimality loss of the different approaches to robust A design for Case study~2.}
\label{fig:perf_rob_DoE_Ades_1stord}
\end{figure}

Overall, we can compare the performance of the different methods for rDoE by a box plot of optimality loss for different true values of the parameters. Figure~\ref{fig:perf_rob_DoE_Ades_1stord} presents the results of 100 Monte Carlo simulations. The central horizontal-line marker indicates the median, the bottom and top edges of the box indicate the $25^\text{th}$ and $75^\text{th}$ percentiles, respectively, the whiskers extend to the most extreme data points not considered outliers, and the outliers are plotted individually using the '+' symbol. The outliers mark the worst-case performance of the compared rDoE approaches.

One can make very similar conclusions at this point as in the previous case study. We summarize these observations in Tab.~\ref{tab:res_perf_1stord}. We rate the approaches as "good", "moderate", or "poor" based on the relative differences between the performance measure in the median, average, and worst cases. Moreover, we use bold font and text underlining to signify best and worst-performing approaches in each of the performance categories, respectively. We can clearly see that the best approaches are sequential (in median performance) and two-stage (in average and worst-case performance) techniques. As in the previous case study, the scenario-based DoE would be rated the best approach if no repeated experiments (with re-estimation) are possible.

\begin{table}
\caption{Comparison of performance of different approaches to rDoE for Case study~2. Bold font and underlined text mark the best and the worst performance, respectively.}
\label{tab:res_perf_1stord}
\centering
\begin{tabular}{c|ccccc}
% \multirow{2}{*}{Performance} & \multicolumn{5}{c}{Approach}\\
% \midrule
\toprule
\diagbox{Performance}{Approach}
& nominal   & sequential & min-max & scenario & two-stage\\
\midrule
median & moderate  & \textbf{good} & \underline{poor} & moderate & moderate \\
average& moderate  & good & \underline{poor} & moderate & \textbf{good}\\
worst-case  & \underline{poor}    & poor & poor & moderate & \textbf{good}\\
\bottomrule
\end{tabular}
\end{table}

\subsection{Case Study 3}
We study a two-parameter reaction-kinetic model of the reaction $\text A \overset{p_1}{\longrightarrow}\text B\overset{p_2}{\longrightarrow}\text{C}$ taken from~\cite{atk68}. The concentration of B follows
\begin{equation}\label{eq:cs_static_model}
 \hat y(\ve p, \tau) = \frac{p_1}{p_1-p_2}\left(\exp(-p_2u_\tau)-\exp(-p_1u_\tau)\right),
\end{equation}
where $u_\tau$ represents the time at which the concentration of component B is measured. Unlike in the previous case study, both the parameters enter the model equation in nonlinear fashion.

The true parameter values are unknown but are assumed to lie within $\ve P:=[0.55, 0.9]\times[0.1, 0.45]$. The nominal values of parameters are taken as the midpoint of $\ve P$. We consider the same measurement error as in the previous case studies. We use again the simplest setup regarding the number of available experiments, taking two measurements in two consecutive experiments---i.e., $N:=4$ and $N_e:=2$.

\begin{table}
\caption{Comparison of nominal A-optimal designs with $N:=4$ for different values of nominal parameters.}
\label{tab:comp_nom_DoE_Ades_reacABC}
\centering
\begin{tabular}{cc|cccc}
\toprule
\multicolumn{2}{c}{Parameter values} & \multicolumn{4}{c}{Design}\\
\midrule
$p_1^*\equiv \hat p_1$ & $p_2^*\equiv \hat p_2$ & $u_1^*$ & $u_2^*$ & $u_3^*$ & $u_4^*$\\
\midrule
0.55 & 0.275 & 1.22 & 1.22 & 1.22 & 6.46\\
0.9  & 0.275 & 0.85 & 0.85 & 0.85 & 5.26\\
0.55 & 0.45  & 1.16 & 1.16 & 5.13 & 5.13\\
\bottomrule
\end{tabular}
\end{table}
    
\begin{figure}
\centering
\psfrag{p1}[cc][cc][0.8]{$p_1$}
\psfrag{p2}[cc][cc][0.8]{$p_2$}
\includegraphics[width=0.8\linewidth]{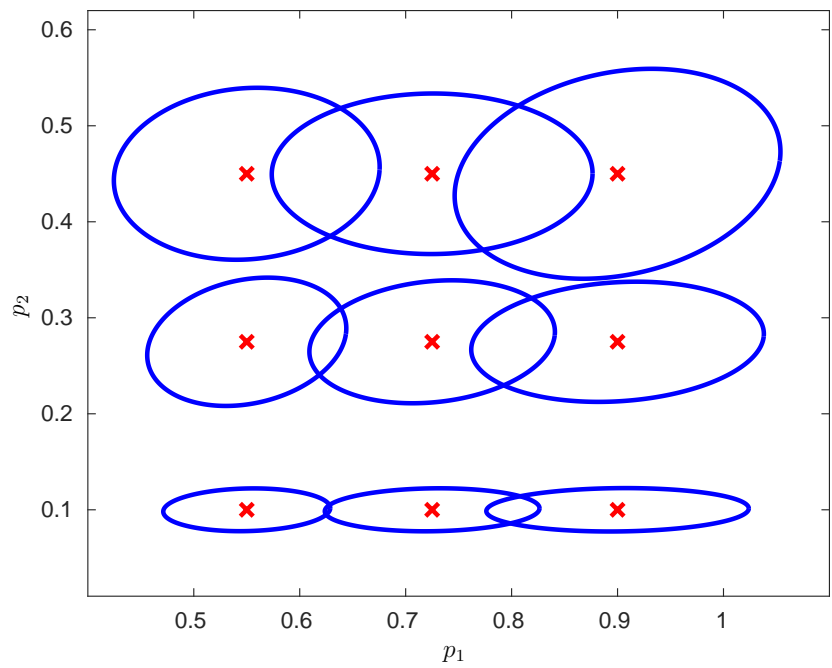}
\caption{Comparison of 2$\sigma$-joint-confidence regions for different values of the parameters for Case study~3.}
\label{fig:conf_reg_comp_reacABC}
\end{figure}

To demonstrate the nature of the case study, we calculate the nominal designs for $N:=4$ corresponding to all combinations of minimal, nominal, and maximal values of parameters within $\ve P$. The general structure of the designs is to repetitively take measurements at $u_\tau\in[0.74, 1.53]$ to better estimate $p_1$ and at $u_\tau\in[3.95, 11.53]$ to better estimate $p_2$. Table~\ref{tab:comp_nom_DoE_Ades_reacABC} shows the nominal A-optimal design for some selected values of nominal parameters. One can clearly observe that the change in nominal parameter value influences both the design points---i.e., in contrast with the previous case study. Secondly, the number of mutual design points varies for different parameter values. This feature appears to be related to the differences between the two parameters---if the parameter values are close to each other, the experiments should distinguish the parameter values more carefully. Figure~\ref{fig:conf_reg_comp_reacABC} shows the corresponding 2$\sigma$-joint-confidence regions. It is clear that higher parameter values result in larger confidence regions---i.e., in harder estimation.

\begin{figure}
\centering
\psfrag{delta}[cc][tt][0.8]{$\phi_A^{\text{seq}}-\phi_A^{\text{2-stg}}$}
\psfrag{p1}[cc][cc][0.8]{$p_1$}
\psfrag{p2}[cc][cc][0.8]{$p_2$}
\psfrag{SEQUENTIDOMINATES}[cc][cc][0.8]{$\!\!\phi_A^{\text{seq}}-\phi_A^{\text{2-stg}} < 0$}
\psfrag{TWOSTAGEDOMINATES}[cc][cc][0.8]{$\!\!\phi_A^{\text{seq}}-\phi_A^{\text{2-stg}} > 0$}
\includegraphics[width=0.8\linewidth]{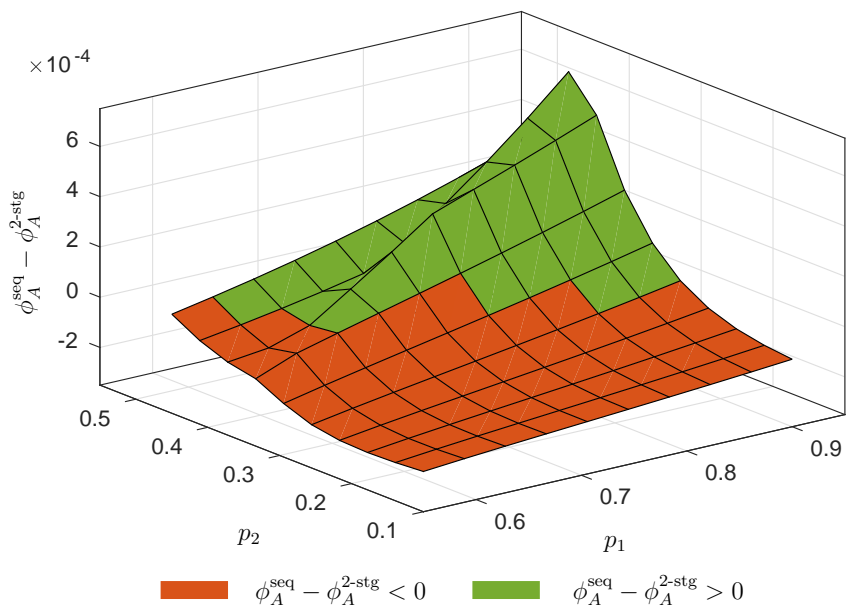}
\caption{Optimality-loss comparison of the sequential ($\phi_A^{\text{seq}}$) and the two-stage ($\phi_A^{\text{2-stg}})$ approaches to robust A design for different true values of parameters for Case study~3.}
\label{fig:perf_seq_2stg_comp_Ades_reacABC}
\end{figure}

The computational setup---regarding the use of BARON and tuning and evaluation of rDoE approaches---is the same as in Case studies~1\&2. As in the previous case study, we first evaluate the performance of the two most promising approaches. Figure~\ref{fig:perf_seq_2stg_comp_Ades_reacABC} shows the corresponding results from Monte Carlo simulation with varied true values of the parameters. The same type of plot is given as in Fig.~\ref{fig:perf_seq_2stg_comp_Ades_1stord}. The conclusions implied by the plot are similar to those observed in Case study~2. The sequential DoE is best in median performance and when the true values of the parameters are close to the nominal ones. Also, the cases, where the performance of the sequential DoE is superior, are related to the instances of true parameter values, where the estimation is more effective---i.e., where the low values of parameters result in large confidence regions.

\begin{figure}
\centering
\psfrag{det}[cc][tt][0.8]{$\phi_A-\phi_A^*$}
\psfrag{strat}[tt][bb][0.8]{$\leftarrow$ DoE strategy $\rightarrow$}
\psfrag{nom}[cc][bb][0.8]{nominal}
\psfrag{seq}[cc][bb][0.8]{sequential}
\psfrag{mmax}[cc][bb][0.8]{min-max}
\psfrag{rob}[cc][bb][0.8]{scenario}
\psfrag{rob2}[cc][bb][0.8]{two-stage}
\includegraphics[width=0.7\linewidth]{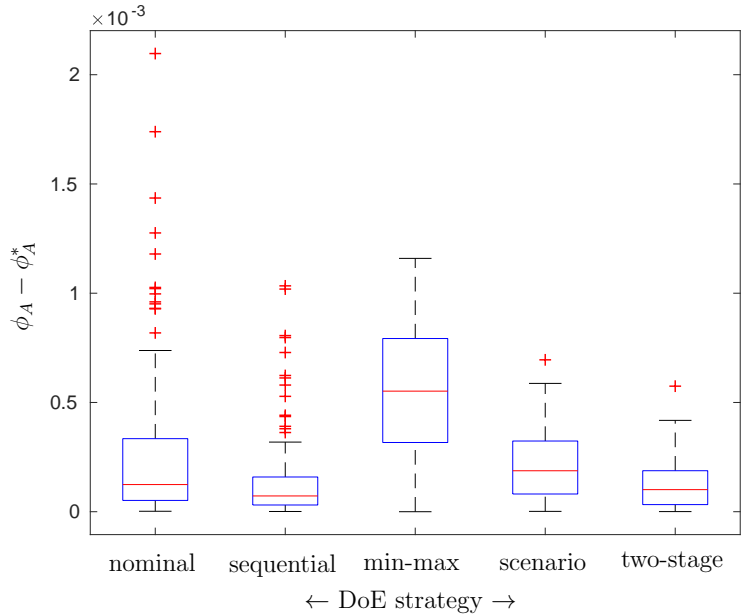}
\caption{A box plot of performance of the different approaches to robust A design for Case study~3.}
\label{fig:perf_rob_DoE_Ades_reacABC}
\end{figure}

Finally, we can compare the performance of the different methods for rDoE measured by the optimality loss of the respective A-optimal design criterion. Figure~\ref{fig:perf_rob_DoE_Ades_reacABC} presents the results of 100 Monte Carlo simulation using box plots. We can generally make similar conclusions as in Case study~2 and in Table~\ref{tab:res_perf_1stord}. The most striking differences are:
\begin{itemize}
  \item The scenario-based design achieves very good performance \wrt to the worst-case performance.
 \item The two-stage design does not improve by much the performance of the scenario-based DoE, which performs extremely well though, yet achieves the best mean and worst-case performance.
\end{itemize}

\subsection{Case Study 4}
In the final case study, we study the case of highly-nonlinear model with four estimated parameters. We use a challenging estimation problem in microbial growth to describe the effect of culture temperature $u_\tau$ on the growth rate of a microbial population $y$. The so-called {\em cardinal temperature model}~\citep{lobr1991} reads as
\begin{equation}
\label{eq:card}
\hat y(\ve p, \tau) = p_1\left[1- \frac{(u_\tau - p_2)^2}{(u_\tau - p_2)^2 + u_\tau(p_3 + p_4 - u_\tau) - p_3p_4} \right],
\end{equation}
% \mu_{\rm opt} T_{\rm opt} T_{\rm min} T_{\rm max}
where $u_\tau\in[288, 333]\,\unit{K}$, $p_1$ is the maximal growth rate, $p_2$ corresponds to the optimal growth temperature, and $p_3$ and $p_4$ represent the minimal and maximal temperatures, respectively.

We assume the true parameter values as $p^* = (1.396\,\unit{h^{-1}}, 313.25\,\unit{K}, 289.40\,\unit{K}, 320.23\,\unit{K})^T$. These are taken from~\cite{per18}, where experimental values from~\cite{rat83} are used for regression. For the designer, the true parameter values are unknown but are assumed to lie within $\ve P:=[1, 2]\,\unit{h^{-1}} \times[300, 320]\,\unit{K}\times [283, 293]\,\unit{K} \times[318, 328]\,\unit{K}$. The nominal values of parameters are taken as the midpoint of $\ve P$. This case study bears certain similarities to the previous examples---such as linear entry of parameter $p_1$ and mutual interactions of the rest of the parameters resulting in varying estimation effectiveness. We consider the measurement error to be a random variable distributed as zero-mean, white Gaussian noise with standard deviation $\sigma = 0.1\,\unit{h}^{-1}$. The setup of the design is again relatively simple regarding the number of available experiments---i.e., $N:=6$ and $N_e:=4$.

To balance the differences in the magnitudes of parameter values, we scale the A-design criterion such that we multiply the first element on the diagonal of the inverse of the Fisher information matrix by a factor of 100. Min-max, scenario-based and two-stage approaches use 81 scenarios with all combinations of minimal, maximal, and nominal values of the parameters given by $\ve P$. To find the different designs, we use Matlab with CasADi~\citep{casadi2019} and Ipopt~\citep{ipopt} as this case study appears numerically challenging for BARON even for the nominal DoE. We also note that---not surprisingly---the problem multi-modality seems to be an arising issue for all the studied designs, which calls for some sophisticated multi-start strategies unless a global solver can be used~\citep{van21}.

The results are very similar in spirit to the previous case studies, evidencing the preceding analyses and the reached conclusions to extend to the more complex problems defined on a larger scale. For demonstration, we evaluate the performance of the different approaches for the case of the aforementioned values of true parameters, which is a representative situation. Taking the optimal DoE as a reference case, percentage of optimality loss is $614\,\%$ (nominal DoE), $179\,\%$ (sequential DoE), $134\,\%$ (min-max DoE), $148\,\%$ (scenario-based DoE), and $94.7\,\%$ (two-stage DoE). Clearly, the scenario-based and two-stage DoE give better results than the sequential approach, which confirms our findings discussed above. The poor performance of the nominal design and the surprisingly good performance of the min-max DoE (the min-max DoE performs better than the scenario-based DoE) can be attributed to the specific choice of the value pair nominal parameters vs. true parameters. This gets revealed in the simulations with varied true parameter values.

% 7.32   34.44   13.46   11.98   11.93    7.55
% loss: 370.4918   83.8798   63.6612   62.9781    3.1421
% imp. over nominal: 60.9175   65.2149   65.3600   78.0778
% ans =
% 
%   280.0000  280.0000  280.0000  280.0000  280.0000  280.0041
%   280.0000  309.9994  307.3235  296.2806  294.0070  289.2205
%   280.0000  295.1602  326.2892  314.3187  311.5084  321.1753
%   321.1435  309.9994  333.0000  318.1014  318.1015  332.5867
%   323.0116  327.1218  319.9153  321.2707  321.2724  323.1969
%   333.0000  333.0000  333.0000  327.3978  327.9719  280.0000

\subsection{Discussion}
Overall, we can say that the more prominent the effect of model nonlinearity on the estimation effectiveness, the more appropriate the advanced robust designs get (scenario-based and multi-stage approaches). The scenario-based DoE appears as a very good strategy for the robust design of experiments for nonlinear parameter estimation. It was shown that the proposed multi-stage approach can improve the performance of the scenario-based DoE and the significance of this improvement appears to be increasing with the range of a priori parametric uncertainty. This approach thus clearly stands for a viable alternative to more standard robustification approaches to DoE. The proposed strategy achieves very good performance in terms of average and worst-case performance. We note here that its performance can be even improved in more advanced rolling-horizon strategies, if one considers the setup with unknown measurement variance, or if the number of the experiments to be designed is large.

The presented case studies show a great potential of rDoE methods in general and scenario-based approaches in particular. One can expect the performance improvement to be even larger for more complex problems as documented in Case study~4. This improvement might vary case-by-case and one would need to conduct some pre-investigation to decide on an appropriate strategy. In this sense, the proposed multi-stage design appears to be an appropriate choice when pre-investigation would be left out.

\begin{figure}
\centering
\psfrag{cpu}[cc][tt][0.8]{CPU time\,[s]}
\psfrag{CS1}[cc][bb][0.8]{Case study~1}
\psfrag{CS2}[cc][bb][0.8]{Case study~2}
\psfrag{CS3}[cc][bb][0.8]{Case study~3}
\psfrag{CS4}[cc][bb][0.8]{Case study~4}
\includegraphics[width=0.7\linewidth]{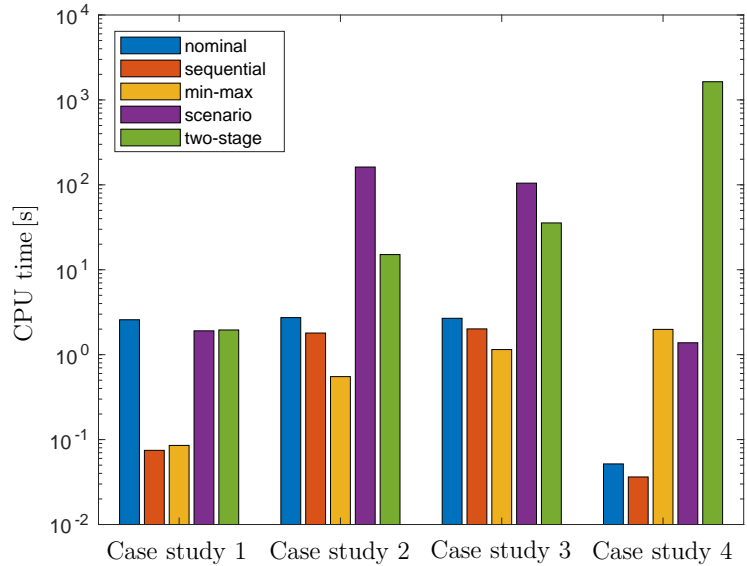}
\caption{CPU time comparison for different DoE approaches over all the case studies.}
\label{fig:perf_rob_DoE_Ades_cpu}
\end{figure}

Let us finally turn the attention towards computational aspects of the presented strategies. The increase in the computational burden of the advanced DoE techniques might be dramatic as far as the complexity of the problem---mostly through the number of parameters---increases.
This is generally confirmed by Figure~\ref{fig:perf_rob_DoE_Ades_cpu} that shows the comparison of average CPU times for computing the studied designs. We compare only computational costs for calculating the initial design---i.e., $N$ experiments for the nominal, min-max, and scenario-based DoE, first $N_e$ experiments for the sequential and two-stage DoE. We note that relatively low CPU times of all the designs but two-stage DoE in the fourth case study are resulting from the use local solver as opposed to the use of BARON in case studies 1--3. BARON is tuned here to heuristic termination by setting the option \verb|DeltaTerm| to the value 1. This makes BARON terminate when no progress is made over several consecutive iterations. Such a setup allows for a fair comparison between the approaches.

Looking at the case studies 1--3, we can interestingly see that the scenario-based strategy is the worst case computationally despite the two-stage DoE involving more optimization variables. Also, the nominal DoE compares badly to the min-max approach despite having the same number of degrees of freedom. We attribute this behavior to the good performance of BARON's domain reduction techniques for small-scale problems. This feature might also be responsible for the lower computational burden of the min-max DoE compared to the nominal and sequential strategies for the case studies 2--3, which is another interesting observation. The computational effort in case study~4---i.e., only solved by a local solver---also supports the former claims.

For the case study~4, we can observe a dramatic increase of CPU time required, considering that a global solver is used in the case studies 2--3, which calls local optimization frequently. The dramatic increase in computational burden is observed regardless of the DoE strategy yet it is most visible in the case of advanced designs (scenario-based and min-max) and, especially, the proposed two-stage design. The local solver apparently faces certain numerical difficulties that arise from the nonlinear and non-convex nature of the problem as well as---for the case of the two-stage design---from an increased number of variables. A remedy to the observed issues might lie in approaches that look into the problem of the computational efficiency of scenario-based optimization and decouple the sparsely structured optimization problems to resolve them efficiently~\citep{mar15, kri19}.

\section{Conclusions}\label{sec:conclusions}
In this paper, we compared some well-known schemes for the robust design of experiments (rDoE). We have also proposed and studied a novel method based on multi-stage decision making, which is found to be a viable methodology for (re-)design of experiments. We used three simple nonlinear case studies and one example on a larger scale for evaluation of the performance of the presented method and we have shown several interesting insights into the problem of rDoE. Based on this study, we conclude that the use of the multi-stage design of experiments becomes more promising with the increasing nonlinearity of the studied system, increasing number of experiments to design, as well as with the increasing range of a priori parametric uncertainty. If, for example, the uncertainty range is narrow, one would use scenario-based formulation, which is computationally less intensive. Among the directions of further research, one could explore the joint robust optimization of the performance mean and variance or use the distributionally robust optimization~\citep{moh18}.

\section*{Acknowledgments}
The authors acknowledge the German Academic Exchange Service (DAAD) and The Ministry of Education, Science, Research and Sport of the Slovak Republic under the Exchange Involving Project ``Reliable and Real-time Feasible Estimation and Control of Chemical Plants''. RP and MF acknowledge the contribution of the Slovak Research and Development Agency (projects APVV 15-0007 and APVV 20-0261), of the Scientific Grant Agency of the Slovak Republic (grant no. 1/0691/21), and of the European Commission (grant no. 790017).

\bibliographystyle{elsarticle-harv}
\bibliography{references}

\begin{thebibliography}{60}
\expandafter\ifx\csname natexlab\endcsname\relax\def\natexlab#1{#1}\fi
\expandafter\ifx\csname url\endcsname\relax
  \def\url#1{\texttt{#1}}\fi
\expandafter\ifx\csname urlprefix\endcsname\relax\def\urlprefix{URL }\fi

\bibitem[{Andersson et~al.(2019)Andersson, Gillis, Horn, Rawlings, and
  Diehl}]{casadi2019}
Andersson, J. A.~E., Gillis, J., Horn, G., Rawlings, J.~B., Diehl, M., 2019.
  {CasADi} -- {A} software framework for nonlinear optimization and optimal
  control. Mathematical Programming Computation 11~(1), 1--36.

\bibitem[{Asprey and Macchietto(2002)}]{asp02}
Asprey, S., Macchietto, S., 2002. Designing robust optimal dynamic experiments.
  Journal of Process Control 12~(4), 545 -- 556.

\bibitem[{Atkinson and Hunter(1968)}]{atk68}
Atkinson, A.~C., Hunter, W.~G., 1968. The design of experiments for parameter
  estimation. Technometrics 10~(2), 271--289.

\bibitem[{Barz et~al.(2010)Barz, Arellano-Garcia, and Wozny}]{bar_seqoed}
Barz, T., Arellano-Garcia, H., Wozny, G., 2010. Handling uncertainty in
  model-based optimal experimental design. Industrial \& Engineering Chemistry
  Research 49~(12), 5702--5713.

\bibitem[{Bates and Watts(1988)}]{bat88}
Bates, D.~M., Watts, D.~G., 1988. Nonlinear Regression Analysis and Its
  Applications. John Wiley \& Sons, Inc.

\bibitem[{Box et~al.(1978)Box, Hunter, Hunter, et~al.}]{box78}
Box, G.~E., Hunter, W.~G., Hunter, J.~S., et~al., 1978. Statistics for
  experimenters. Vol. 664. John Wiley and sons New York.

\bibitem[{Campi and Weyer(2005)}]{cam05}
Campi, M., Weyer, E., 2005. Guaranteed non-asymptotic confidence regions in
  system identification. Automatica 41~(10), 1751 -- 1764.

\bibitem[{Espie and Macchietto(1989)}]{esp89}
Espie, D., Macchietto, S., 1989. The optimal design of dynamic experiments.
  AIChE Journal 35~(2), 223--229.

\bibitem[{Fedorov(2013)}]{fed13}
Fedorov, V.~V., 2013. Theory of optimal experiments. Elsevier.

\bibitem[{Filatov and Unbehauen(2004)}]{filatov04}
Filatov, N., Unbehauen, H., 2004. Adaptive Dual Control. Lecture Notes in
  Control and Information Sciences. Springer Verlag, Berlin.

\bibitem[{Fisher(1935)}]{fis35}
Fisher, R., 1935. The Design of Experiments. Oliver \& Boyd.

\bibitem[{Franceschini and Macchietto(2008)}]{fra08}
Franceschini, G., Macchietto, S., 2008. Model-based design of experiments for
  parameter precision: State of the art. Chem. Eng. Sci. 63~(19), 4846--4872.

\bibitem[{Fung et~al.(2016)Fung, Ng, Zhang, and Gani}]{fun16}
Fung, K.~Y., Ng, K.~M., Zhang, L., Gani, R., 2016. A grand model for chemical
  product design. Computers \& Chemical Engineering.

\bibitem[{Galvanin et~al.(2006)Galvanin, Barolo, Bezzo, and Macchietto}]{gal06}
Galvanin, F., Barolo, M., Bezzo, F., Macchietto, S., 2006. A framework for
  model-based design of parallel experiments in dynamic systems. In: Marquardt,
  W., Pantelides, C. (Eds.), 16th European Symposium on Computer Aided Process
  Engineering and 9th International Symposium on Process Systems Engineering.
  Vol.~21 of Computer Aided Chemical Engineering. Elsevier, pp. 249 -- 254.

\bibitem[{Galvanin et~al.(2010)Galvanin, Barolo, Bezzo, and Macchietto}]{gal10}
Galvanin, F., Barolo, M., Bezzo, F., Macchietto, S., 2010. A backoff strategy
  for model-based experiment design under parametric uncertainty. AIChE Journal
  56~(8), 2088--2102.

\bibitem[{Garstka and Wets(1974)}]{Garstka1974}
Garstka, S.~J., Wets, R. J.-B., 1974. On decision rules in stochastic
  programming. Math Program 7~(1), 117--143.

\bibitem[{Georgakis(2013)}]{geo13}
Georgakis, C., 2013. Design of dynamic experiments: A data-driven methodology
  for the optimization of time-varying processes. Industrial \& Engineering
  Chemistry Research 52~(35), 12369--12382.

\bibitem[{Goodwin and Payne(1977)}]{goodwin1977dynamic}
Goodwin, G., Payne, R., 1977. Dynamic system identification. experiment design
  and data analysis.

\bibitem[{{Gottu Mukkula} and Paulen(2017)}]{anwesh17esc}
{Gottu Mukkula}, A.~R., Paulen, R., 2017. Robust model-based design of
  experiments for guaranteed parameter estimation. In: Espuña, A., Graells,
  M., Puigjaner, L. (Eds.), 27th European Symposium on Computer Aided Process
  Engineering. Vol.~40 of Computer Aided Chemical Engineering. Elsevier, pp.
  1639 -- 1644.

\bibitem[{{Gottu Mukkula} and Paulen(2019)}]{mupa19_jpc}
{Gottu Mukkula}, A.~R., Paulen, R., 2019. Optimal experiment design in
  nonlinear parameter estimation with exact confidence regions. Journal of
  Process Control 83, 187 -- 195.

\bibitem[{Hjalmarsson(2005)}]{hja05}
Hjalmarsson, H., 2005. From experiment design to closed-loop control.
  Automatica 41~(3), 393--438.

\bibitem[{Holtorf et~al.(2019)Holtorf, Mitsos, and Biegler}]{hol19}
Holtorf, F., Mitsos, A., Biegler, L.~T., 2019. Multistage nmpc with on-line
  generated scenario trees: Application to a semi-batch polymerization process.
  Journal of Process Control 80, 167 -- 179.

\bibitem[{K\"orkel et~al.(2004)K\"orkel, Kostina, Bock, and Schl\"oder}]{kor04}
K\"orkel, S., Kostina, E., Bock, H., Schl\"oder, J., 2004. Numerical methods
  for optimal control problems in design of robust optimal experiments for
  nonlinear dynamic processes. Optimization Methods and Software 19~(3-4),
  327--338.

\bibitem[{Krishnamoorthy et~al.(2019)Krishnamoorthy, Foss, and
  Skogestad}]{kri19}
Krishnamoorthy, D., Foss, B., Skogestad, S., 2019. A primal decomposition
  algorithm for distributed multistage scenario model predictive control.
  Journal of Process Control 81, 162 -- 171.

\bibitem[{Lobry et~al.(1991)Lobry, Rosso, and Flandrois}]{lobr1991}
Lobry, J., Rosso, L., Flandrois, J., 1991. A fortran subroutine for the
  determination of parameter confidence limits in non-linear models. Binary 3,
  86--93.

\bibitem[{Lucia et~al.(2013)Lucia, Finkler, and Engell}]{lucia2013}
Lucia, S., Finkler, T., Engell, S., 2013. Multi-stage nonlinear model
  predictive control applied to a semi-batch polymerization reactor under
  uncertainty. Journal of Process Control 23, 1306--1319.

\bibitem[{M{\aa}rtensson and Hjalmarsson(2006)}]{mar06}
M{\aa}rtensson, J., Hjalmarsson, H., 2006. Robust input design using sum of
  squares constraints. IFAC Proceedings Volumes 39~(1), 1352 -- 1357, 14th IFAC
  Symposium on Identification and System Parameter Estimation.

\bibitem[{Mart{\'{i}} et~al.(2015)Mart{\'{i}}, Lucia, Sarabia, Paulen, Engell,
  and {de Prada}}]{mar15}
Mart{\'{i}}, R., Lucia, S., Sarabia, D., Paulen, R., Engell, S., {de Prada},
  C., 2015. Improving scenario decomposition algorithms for robust nonlinear
  model predictive control. Computers \& Chemical Engineering 79, 30 -- 45.

\bibitem[{Mesbah and Streif(2015)}]{mes15}
Mesbah, A., Streif, S., 2015. A probabilistic approach to robust optimal
  experiment design with chance constraints. IFAC-PapersOnLine 48~(8), 100 --
  105, 9th IFAC Symposium on Advanced Control of Chemical Processes ADCHEM
  2015.

\bibitem[{Mohajerin~Esfahani and Kuhn(2018)}]{moh18}
Mohajerin~Esfahani, P., Kuhn, D., 2018. Data-driven distributionally robust
  optimization using the wasserstein metric: performance guarantees and
  tractable reformulations. Mathematical Programming 171, 115 -- 166.

\bibitem[{Nimmegeers et~al.(2020)Nimmegeers, Bhonsale, Telen, and {Van
  Impe}}]{nim20}
Nimmegeers, P., Bhonsale, S., Telen, D., {Van Impe}, J., 2020. Optimal
  experiment design under parametric uncertainty: A comparison of a
  sensitivities based approach versus a polynomial chaos based stochastic
  approach. Chemical Engineering Science 221, 115651.

\bibitem[{Olofsson et~al.(2019)Olofsson, Hebing, Niedenführ, Deisenroth, and
  Misener}]{olo19}
Olofsson, S., Hebing, L., Niedenführ, S., Deisenroth, M.~P., Misener, R.,
  2019. Gpdoemd: A python package for design of experiments for model
  discrimination. Computers \& Chemical Engineering 125, 54 -- 70.

\bibitem[{Pantelides and Renfro(2013)}]{pan13}
Pantelides, C., Renfro, J., 2013. The online use of first-principles models in
  process operations: Review, current status and future needs. Computers \&
  Chemical Engineering 51, 136--148.

\bibitem[{Paulson et~al.(2019)Paulson, Martin-Casas, and Mesbah}]{pau19}
Paulson, J.~A., Martin-Casas, M., Mesbah, A., 2019. Optimal bayesian experiment
  design for nonlinear dynamic systems with chance constraints. Journal of
  Process Control 77, 155 -- 171.

\bibitem[{Perić et~al.(2018)Perić, Paulen, Villanueva, and Chachuat}]{per18}
Perić, N.~D., Paulen, R., Villanueva, M.~E., Chachuat, B., 2018.
  Set-membership nonlinear regression approach to parameter estimation. Journal
  of Process Control 70, 80 -- 95.

\bibitem[{Petsagkourakis and Galvanin(2021)}]{pet20}
Petsagkourakis, P., Galvanin, F., 2021. Safe model-based design of experiments
  using gaussian processes. Computers \& Chemical Engineering 151, 107339.

\bibitem[{Pronzato(2008)}]{pro08}
Pronzato, L., 2008. Survey paper: Optimal experimental design and some related
  control problems. Automatica 44~(2), 303--325.

\bibitem[{Pronzato and P\'azman(2013)}]{pro13}
Pronzato, L., P\'azman, A., 2013. Design of Experiments in Nonlinear Models:
  Asymptotic Normality, Optimality Criteria and Small-Sample Properties.
  Springer.

\bibitem[{Pronzato and Walter(1985)}]{pronzato1985robust}
Pronzato, L., Walter, E., 1985. Robust experiment design via stochastic
  approximation. Math Biosci 75~(1), 103--120.

\bibitem[{Pronzato and Walter(1988)}]{pronzato1988robust}
Pronzato, L., Walter, E., 1988. Robust experiment design via maximin
  optimization. Math Biosci 89~(2), 161--176.

\bibitem[{Ratkowsky et~al.(1983)Ratkowsky, Lowry, McMeekin, Stokes, and
  Chandler}]{rat83}
Ratkowsky, D., Lowry, R., McMeekin, T., Stokes, A., Chandler, R., 1983. Model
  for bacterial culture growth rate throughout the entire biokinetic
  temperature range. Journal of Bacteriology 154, 1222--1226.

\bibitem[{Rojas et~al.(2007)Rojas, Welsh, Goodwin, and Feuer}]{roj07}
Rojas, C.~R., Welsh, J.~S., Goodwin, G.~C., Feuer, A., 2007. Robust optimal
  experiment design for system identification. Automatica 43~(6), 993 -- 1008.

\bibitem[{Rooney and Biegler(2001)}]{rooney2001}
Rooney, W.~C., Biegler, L.~T., 2001. Design for model parameter uncertainty
  using nonlinear confidence regions. AIChE Journal 47~(8), 1794--1804.

\bibitem[{Safdarnejad et~al.(2016)Safdarnejad, Gallacher, and
  Hedengren}]{saf16}
Safdarnejad, S.~M., Gallacher, J.~R., Hedengren, J.~D., 2016. Dynamic parameter
  estimation and optimization for batch distillation. Computers \& Chemical
  Engineering 86, 18--32.

\bibitem[{Schenkendorf et~al.(2009)Schenkendorf, Kremling, and Mangold}]{sch09}
Schenkendorf, R., Kremling, A., Mangold, M., 02 2009. Optimal experimental
  design with the sigma point method. IET systems biology 3, 10--23.

\bibitem[{Sebastiani and Wynn(2000)}]{seb00}
Sebastiani, P., Wynn, H.~P., 2000. Maximum entropy sampling and optimal
  bayesian experimental design. Journal of the Royal Statistical Society:
  Series B (Statistical Methodology) 62~(1), 145--157.

\bibitem[{Seber and Wild(2003)}]{SeberWild200309}
Seber, G. A.~F., Wild, C.~J., 9 2003. Nonlinear Regression. Wiley-Interscience.

\bibitem[{Simon(1989)}]{sim89}
Simon, R., 1989. Optimal two-stage designs for phase ii clinical trials.
  Controlled Clinical Trials 10~(1), 1 -- 10.

\bibitem[{Streif et~al.(2014)Streif, Petzke, Mesbah, Findeisen, and
  Braatz}]{str14}
Streif, S., Petzke, F., Mesbah, A., Findeisen, R., Braatz, R.~D., 2014. Optimal
  experimental design for probabilistic model discrimination using polynomial
  chaos. 19th {IFAC} World Congress 47~(3), 4103--4109.

\bibitem[{Sverdlov et~al.(2019)Sverdlov, Ryeznik, and Wong}]{sve19}
Sverdlov, O., Ryeznik, Y., Wong, W.~K., 2019. On optimal designs for clinical
  trials: An updated review. Journal of Statistical Theory and Practice
  14~(10), 1 -- 10.

\bibitem[{Tawarmalani and Sahinidis(2005)}]{baron}
Tawarmalani, M., Sahinidis, N.~V., 2005. {A polyhedral branch-and-cut approach
  to global optimization}. Mathematical Programming 103, 225--249.

\bibitem[{Telen et~al.(2014)Telen, Vercammen, Logist, and Impe}]{Telen_roed}
Telen, D., Vercammen, D., Logist, F., Impe, J.~V., 2014. Robustifying optimal
  experiment design for nonlinear, dynamic (bio)chemical systems. Computers \&
  Chemical Engineering 71, 415 -- 425.

\bibitem[{Thangavel et~al.(2018)Thangavel, Lucia, Paulen, and
  Engell}]{Thangavel2018c}
Thangavel, S., Lucia, S., Paulen, R., Engell, S., 2018. Dual robust nonlinear
  model predictive control: A multi-stage approach. Journal of Process Control
  72, 39 -- 51.

\bibitem[{Vanaret et~al.(2021)Vanaret, Seufert, Schwientek, Karpov, Ryzhakov,
  Oseledets, Asprion, and Bortz}]{van21}
Vanaret, C., Seufert, P., Schwientek, J., Karpov, G., Ryzhakov, G., Oseledets,
  I., Asprion, N., Bortz, M., 2021. Two-phase approaches to optimal model-based
  design of experiments: how many experiments and which ones? Computers \&
  Chemical Engineering 146, 107218.

\bibitem[{W{\"a}chter and Biegler(2006)}]{ipopt}
W{\"a}chter, A., Biegler, T.~L., 2006. On the implementation of an
  interior-point filter line-search algorithm for large-scale nonlinear
  programming. Mathematical Programming 106~(1), 25--57.

\bibitem[{Walter and Pronzato(1987)}]{walter1987optimal}
Walter, E., Pronzato, L., 1987. Optimal experiment design for nonlinear models
  subject to large prior uncertainties. American Journal of
  Physiology-Regulatory, Integrative and Comparative Physiology 253~(3),
  R530--R534.

\bibitem[{Walz et~al.(2018)Walz, Djelassi, Caspari, and Mitsos}]{WALZ201892}
Walz, O., Djelassi, H., Caspari, A., Mitsos, A., 2018. Bounded-error optimal
  experimental design via global solution of constrained min-max program.
  Computers \& Chemical Engineering 111, 92--101.

\bibitem[{Walz et~al.(2020)Walz, Djelassi, and Mitsos}]{wal20}
Walz, O., Djelassi, H., Mitsos, A., 2020. Optimal experimental design for
  optimal process design: A trilevel optimization formulation. AIChE Journal
  66~(1), e16788.

\bibitem[{Welsh and Rojas(2009)}]{wel09}
Welsh, J.~S., Rojas, C.~R., 2009. A scenario based approach to robust
  experiment design. IFAC Proceedings Volumes 42~(10), 186 -- 191, 15th IFAC
  Symposium on System Identification.

\bibitem[{Wolkenhauer et~al.(2008)Wolkenhauer, Wellstead, Cho, Banga, and
  Balsa-Canto}]{wol08}
Wolkenhauer, O., Wellstead, P., Cho, K.-H., Banga, J.~R., Balsa-Canto, E., 09
  2008. {Parameter estimation and optimal experimental design}. Essays in
  Biochemistry 45, 195--210.

\end{thebibliography}
\end{document}